\documentclass[a4paper,11pt]{article}
\pdfoutput=1 

\usepackage{jheppub} 
\usepackage{hyperref}
\usepackage[T1]{fontenc} 
\usepackage[normalem]{ulem}

\usepackage{graphicx}
\usepackage{caption}
\usepackage{subcaption}
\usepackage{dcolumn}
\usepackage{bm}

\usepackage{color,xcolor,url,ulem,verbatim,appendix}
\usepackage[colorlinks=true
,urlcolor=blue
,anchorcolor=blue
,citecolor=blue 
,filecolor=blue
,linkcolor=blue
,menucolor=blue
,linktocpage=true
,pdfproducer=medialab
,pdfa=true
]{hyperref}

\newcommand{\beq}{\begin{equation}}
\newcommand{\eeq}{\end{equation}}
\newcommand{\bea}{\begin{eqnarray}}
\newcommand{\eea}{\end{eqnarray}}

\newcommand{\ep}{\epsilon}

\newcommand{\mpl}{M_{\text{Pl}}}

\pdfoutput=1 

\title{\begin{center}
    \LARGE{\textbf{Phase Transitions from the Fifth Dimension}}
\end{center}
}
\author[a]{Kaustubh Agashe}
\emailAdd{kagashe@umd.edu}
\author[b]{Peizhi Du}
\emailAdd{peizhi.du@stonybrook.edu}
\author[a]{Majid Ekhterachian}
\emailAdd{ekhtera@umd.edu}
\author[a,c,d]{Soubhik Kumar}
\emailAdd{soubhik@berkeley.edu}
\author[a]{Raman Sundrum}
\emailAdd{raman@umd.edu}
\affiliation[a]{Maryland Center for Fundamental Physics, Department of Physics,\\ University of Maryland, College Park, MD 20742 USA}
\affiliation[b]{C.N. Yang Institute for Theoretical Physics, Stony Brook University, Stony Brook, NY, 11794, USA}
\affiliation[c]{Berkeley Center for Theoretical Physics, Department of Physics,
University of California, Berkeley, CA 94720, USA}
\affiliation[d]{Theoretical Physics Group, Lawrence Berkeley National Laboratory, Berkeley, CA 94720, USA}

\abstract{We study the cosmological transition of 5D warped compactifications, from the 
high-temperature black-brane phase to the low-temperature Randall-Sundrum I phase.
The transition proceeds via percolation of bubbles of IR-brane nucleating from the black-brane horizon.  The violent bubble dynamics can be a powerful source of observable stochastic gravitational waves.
While bubble nucleation is non-perturbative in 5D gravity, it is amenable to semiclassical treatment in terms of a ``bounce” configuration interpolating between the two phases. 
We demonstrate how such a bounce configuration can be smooth enough to maintain 5D effective field  theory control, and how a simple ansatz for it
places a rigorous lower-bound on the transition rate in the thin-wall regime, and gives plausible estimates more generally. 
When applied to the Hierarchy Problem,  the minimal Goldberger-Wise stabilization of the warped throat leads to a  slow transition with significant supercooling. We demonstrate that a simple generalization of the Goldberger-Wise potential modifies the IR-brane dynamics so that the transition completes more promptly.  
Supercooling determines the dilution of any (dark) matter abundances generated before the transition, potentially at odds with data, while the prompter transition  resolves such tensions. 
We discuss the impact of the different possibilities on the strength of the gravitational wave signals. 
Via AdS/CFT duality  the warped transition gives a theoretically tractable holographic description of the 4D Composite Higgs (de)confinement transition. Our generalization of the Goldberger-Wise mechanism is dual to, and concretely models, our earlier proposal in which the composite dynamics is governed by separate UV and IR RG fixed points.
The smooth 5D bounce configuration we introduce complements the 4D dilaton/radion dominance derivation presented in our earlier work.
}

\begin{document}
\hspace{22em} UMD-PP-020-5, YITP-SB-2020-29
\maketitle
\flushbottom
\section{Introduction}

As the very early universe evolved, it  likely underwent a series of 
phase transitions (PT), during which the degrees of freedom were rearranged or their properties were altered. These cosmological PTs would have had even more dramatic effects if they were first order, in which case they may have generated new matter asymmetries, or 
significantly altered preexisting abundances.
Moreover they would have sourced a background of stochastic gravitational waves (GW) that survive until today, which if observed can give us unique insights into cosmic history \cite{Kosowsky:1991ua,Kosowsky:1992vn,Kosowsky:1992rz,Kamionkowski:1993fg} (see e.g. refs.~\cite{Caprini:2015zlo,Caprini:2019egz} for reviews). 
In the Standard Model (SM), the early universe electroweak (EW) and QCD PTs are not first order (see refs.~\cite{Quiros:1999jp,Petreczky:2012rq} for  review), but  beyond-SM (BSM) extensions or analogs may well be so. BSM scenarios that introduce new first order PTs tied to the EW scale or related to it by naturalness (within a few orders of magnitude) are particularly exciting since they have the potential to both be probed by future gravitational wave experiments \cite{Audley:2017drz,Harry:2006fi,Graham:2017pmn,Kawamura:2011zz,Gong:2014mca} and by future collider experiments.

Composite Higgs theories (see refs.~\cite{Bellazzini:2014yua,Panico:2015jxa} for review) are among the most attractive candidates for BSM physics as they can naturally explain the large observed particle physics hierarchies. In these theories the Higgs boson is a composite of strongly interacting constituents, confined at zero temperature but deconfined at high enough temperatures. Studying this confinement-deconfinemet PT in the cosmological context is thus extremely well-motivated.
However the confining dynamics is necessarily non-perturbatively strong, making it very challenging to analyze.

These theories often need large number of degrees of freedom $\sim N^2$ of the confined constituents and an approximate conformal symmetry in order to generate the large flavor hierarchies.
But this conformal symmetry must break down at the confinement scale. In an earlier paper \cite{Agashe:2019lhy} we focused on the scenario  where the breakdown of conformal invariance is \emph{spontaneous}, which we dubbed ``spontaneous confinement''.
 There we clarified the parametric regime where the dominant piece of bounce action, controlling the transition rate, can be computed in the confined phase and within the effective field theory (EFT) of the pseudo Nambu-Goldstone boson (pNGB) of spontaneously broken conformal invariance, namely the dilaton \cite{Salam:1970qk,Rattazzi:2000hs,Goldberger:2008zz,Chacko:2012sy,Bellazzini:2012vz,Chacko:2013dra,Chacko:2014pqa}. This parametrically dominant piece, considered in the pioneering ref.~\cite{Creminelli:2001th}, accounts only for the approach of the confined phase towards deconfinement, missing the final transition to deconfinement itself where the degrees of freedom are rearranged and the dilaton EFT breaks down.
 Missing the physics of this final transition to deconfinement means that our final approximate transition rates were useful but still crude.

Further theoretical control is possible when a holographic AdS/CFT \cite{Maldacena:1997re,Gubser:1998bc,Witten:1998qj,Aharony:1999ti} dual can be formulated. The dual description involves warped compactification of extra dimension(s) as in Randall-Sundrum (RS) models \cite{Randall:1999ee,Randall:1999vf},\cite{ArkaniHamed:2000ds,Rattazzi:2000hs} (see \cite{Davoudiasl:2009cd,Gherghetta:2010cj} for reviews). These theories exhibit  a finite temperature black-brane solution as well as a low (or zero) temperature standard RS1 solution. The transition between these two phases is an analog of the well-known Hawking-Page PT of global AdS \cite{Hawking:1982dh}, but in the AdS Poincare patch relevant here, there are significant differences. The transition considered here is dual to the (de)confinement PT mentioned above, where the black brane is dual to the deconfined phase and RS1 is dual to the confined phase.
Remarkably, in this dual RS1 formulation, the transition is a non-perturbative 5D quantum gravity process ($\sim e^{-1/G_{N,5D}}$) in which bubbles of IR-brane nucleate from the black-brane horizon, expand and collide, producing an observable stochastic gravitational wave background! And remarkably again, this non-perturbative effect can be captured by semiclassical methods.

This higher dimensional description has been used to study this PT \cite{Creminelli:2001th,Randall:2006py, Nardini:2007me, Konstandin:2010cd, Konstandin:2011dr,Megias:2018sxv,Megias:2020vek,Fujikura:2019oyi, Bigazzi:2020phm}, using an ansatz for the bounce configuration describing critical bubbles first introduced in \cite{Creminelli:2001th}. One might have hoped that the 5D EFT is controlled  despite the inevitable breakdown of the 4D dilaton EFT described above.
However this particular ansatz is not controlled,
even in the 5D EFT, as will be discussed in section \ref{sec:bounce}. 
Quite apart from the detailed extremization of the bounce configuration, 5D EFT control hinges on a qualitative puzzle: how to smoothly interpolate in space-time between the IR-brane and black-brane phases? 
In this paper, we begin by solving this qualitative problem and show what conditions it places on possible bounce configurations. However, even within this smooth class it is technically challenging to find the extremized bounce configuration that dominates the transition rate. 
Instead, we introduce a new bounce ansatz within this smooth controlled class, depicted in figure \ref{figure:Ansatz_topology}, and show that it gives a rigorous and useful lower bound on the transition rate in the thin-wall regime and a very plausible estimate of the rate more generally.  

In principle, the smooth class of bounce configurations can be extremized with respect to the 5D action to determine the true transition rate, rather than settling for an ansatz which at best bounds this rate. While we are currently unable to accomplish this feat, it should be noted that in a roughly analogous 6D EFT a domain wall solution between a black-brane phase and an ``IR-brane'' phase was derived, exploiting a $Z_2$ symmetry between  thermal Euclidean ``time'' and the sixth dimension \cite{Aharony:2005bm}. In the thin-wall limit this solution can be recast as a bounce solution for transitioning between the two phases, which we hope to explore in future work.

Both our ansatz and the work related to the prior ansatz of \cite{Creminelli:2001th} share the correct parametrically dominant dilaton contribution to the bounce. Our improvement is at the level of the parametrically subdominant
corrections involving the IR-brane/black-brane juncture. However, these corrections are qualitatively 
significant as discussed above, and quantitatively significant for realistic choices of parameters. 

In composite Higgs models, the large hierarchy between the Planck and the weak scales is explained by 
a small deviation of the theory from scale invariance, where the parameter characterizing the small deviation generates the weak scale by dimensional transmutation.
In the minimal models, the same small parameter however significantly suppress the transition rate, forbidding the completion of PT or delaying it until after a large amount of supercooling. For further studies along these lines see refs.~\cite{Konstandin:2011dr,vonHarling:2017yew,Bruggisser:2018mrt,Bruggisser:2018mus,Baratella:2018pxi}. Large supercooling would strongly dilute any primordial (dark) matter abundances, produced before the PT, potentially invalidating such primordial production as the dominant source of (dark) matter seen today.
In \cite{Agashe:2019lhy} we showed that in a scenario where the composite theory runs from the proximity of an ultraviolet (UV) renormalization group (RG) fixed point (FP) to that of an infrared (IR) FP, it is possible to have a much faster transition rate, avoiding large primordial matter dilution. In this scenario the small anomalous dimension corresponding to the UV FP generates the large hierarchy, while the anomalous dimension corresponding to the IR FP, which can be much larger, controls the transition rate. In the present paper, we give a robust AdS/CFT dual realization of this two FP scenario in an explicit 5D model, where the extrema of a generic potential for the Goldberger-Wise scalar field \cite{Goldberger:1999uk} play the role of the FPs. Refs.~\cite{Hassanain:2007js,Konstandin:2010cd,Dillon:2017ctw,Bunk:2017fic,Megias:2018sxv,Fujikura:2019oyi} also explored the possibility of prompter PTs within other non-minimal models.

Using our ansatz we estimate the bounce action, which controls the transition rate, for the minimal model and for the two-FP scenario. Comparing this rate with the dilaton/radion dominance approximation, we identify the subleading corrections and see that in the paramateric regime identified for dilaton dominance the two approaches agree, as expected.
Then by comparing the transition rate with the rate of expansion of the universe, we determine if the transition completes, and when it does, we find the temperature at which this happens. 
We find that staying in the regime where there is a controlled semiclassical approximation to the rate, and for a realistic region of parameter space for which the PT completes, the corrections to the 4D dilaton dominance approximation are quantitatively important, but are captured by our 5D EFT treatment. 

As pointed out in ref.~\cite{Randall:2006py,Konstandin:2011dr} and systematically derived in \cite{Agashe:2019lhy}, the strength of the  stochastic gravitational wave signal arising from the PT (specifically from the better understood bubble collisions) is correlated with the degree of supercooling (and matter dilution). We quantitatively present this relationship using our 5D results and the variability of supercooling  in our two-FP scenario. Of course, the strength of the stochastic gravitational wave signal is critical for being able to detect beyond astrophysical background and detector noise, but the signals can be large enough that even the primordial anisotropies (analogous to those famously seen in the cosmic microwave background) could be observable at future detectors \cite{Geller:2018mwu}.
The gravitational wave detectors have to have sensitivity at frequencies determined by the critical temperature $T_c$ of the PT. In the composite Higgs scenario $T_c\sim \mathcal{O}$(TeV), corresponding to $\mathcal{O}$(mHz) detection frequencies, but quite different frequencies and $T_c$ are possible if an analogous PT occurs in a hidden sector \cite{Schwaller:2015tja}. The results of our paper are straightforwardly transferable to such hidden sector PTs.

Rest of the paper is organized as follows. In section \ref{sec:equilibrium}, we describe the two phases in equilibrium, and obtain the critical temperature for the PT. In section \ref{sec:bounce} we show how we can smoothly interpolate between the two phases and present our ansatz for the bounce. In section \ref{sec:thinwall} we analyze the PT in the thin-wall regime. In section~\ref{sec:twoFP}, we review our two-FP scenario in which the transition rate can be significantly enhanced and present a robust 5D model realizing this scenario. In section \ref{sec:PT_not_thin_wall}, we analyze the PT in the supercooled regime, obtain the temperature for which the PT completes, and discuss the implications for (dark) matter genesis. We discuss the gravitational wave signal produced by the PT in section \ref{sec:GrWaves}, and our conclusions in section \ref{sec:conclusions}. Throughout this paper, we will use the mostly plus i.e. $(-,+,...)$ convention for the Lorentzian metrics.

\section{Equilibrium description of the two phases} \label{sec:equilibrium}
We start with the general 5D action $S_{\text{5D}}$, which is a sum of $S_{\text{GR}}$, the gravitational action and $S_{\chi}$, the action of Goldberger-Wise field $\chi$:
\begin{align} \label{eq:RSaction}
S_{\text{5D}} &= S_{\text{GR}}+ S_{\chi} \nonumber \\
&= 2M_5^3\int d^5x\sqrt{-g}\left(R_5+12 k^2\right)+4M_5^3 k\int d^4x\sqrt{-\gamma}K-\tau_{\text{bd}}\int d^4x\sqrt{-\gamma}+S_{\chi}.
\end{align}
Here $M_5,k$ are the 5D Planck scale and the AdS curvature scale, respectively. Along with a bulk term that contains the 5D Ricci scalar $R_5$, $S_{\text{GR}}$ also contains a Gibbons-Hawking-York boundary term \cite{PhysRevLett.28.1082,PhysRevD.15.2752} which make the variation of the bulk action well defined in the presence of boundaries. This boundary term is characterized by $K=g^{\mu\nu}K_{\mu\nu}$, the trace of the extrinsic curvature $K_{\mu\nu}$ of the boundary and, the induced metric $\gamma_{\mu\nu}$ on the boundary. Lastly, $\tau_{\text{bd}}$ denotes the tension on the boundary. In the following we will always work in units where $k=1$, unless explicitly mentioned.

We first start by neglecting the contribution due to $S_\chi$. In that case the RS metric \cite{Randall:1999ee,Randall:1999vf},
\begin{equation}\label{eq:RSmetric}
ds^2 = -\rho^2 dt^2+\rho^2 \sum_i dx_i^2+\frac{d\rho^2}{\rho^2},
\end{equation}
with the extra dimensional coordinate $\rho$ ranging between $\Lambda_{\text{IR}}<\rho < \Lambda_{\text{UV}}$, represents a solution to $S_{\text{GR}}$. Here $\Lambda_{\text{IR(UV)}}$ represent the location of the IR (UV) boundary. The tensions on the UV and the IR boundaries are given by $+12M_5^3$ and $-12M_5^3$ respectively. Given a choice of $\Lambda_{\text{UV}}$, the scale $\Lambda_{\text{IR}}$
corresponds to the VEV of the ``radion'' field $\phi(x)$ that characterizes the dynamical size of the extra dimension. At this stage $\Lambda_{\text{IR}}$ is an arbitrary integration constant corresponding to the fact that
$\phi$ is a flat direction with no potential. In the dual 4D theory, $\Lambda_{\text{IR}}\equiv \langle\phi\rangle$ spontaneously breaks the (approximate) conformal symmetry of the composite Higgs dynamics and corresponds to the resulting \textit{spontaneous confinement} scale. The associated Goldstone boson is the dilaton, dual to the radion $\phi(x)$ \cite{Rattazzi:2000hs,ArkaniHamed:2000ds}.

To have a predictive theory of $\Lambda_{\text{IR}}$ and avoid a massless radion $\phi$, the Goldberger-Wise
action $S_\chi$ needs to be included as a weak perturbation. This leads to a potential for $\phi$ naturally yielding a hierarchical separation between $\Lambda_{\text{IR}}$ and $\Lambda_{\text{UV}}$.
Over the stabilized hierarchy the RS metric then models the confined phase at $T=0$. Before discussing the details of the Goldberger-Wise stabilization, let us model the deconfined phase at high $T$.

At high temperature, $T\gg \Lambda_{\textrm{IR}}$, but $T\ll \Lambda_{\textrm{UV}}$, there is another approximate solution to $S_{\text{GR}}$ in eq.~\eqref{eq:RSaction} given by the AdS-Schwarzschild (AdS-S) metric,
\begin{equation}\label{eq:AdSSmetric}
ds^2 = -\left(\rho^2-\frac{\rho_h^4 }{\rho^2}\right)dt^2+\rho^2\sum_i dx_i^2+\frac{d\rho^2}{\rho^2 -\frac{\rho_h^4}{\rho^2}}.
\end{equation}
This solution is exact for $\Lambda_{\textrm{UV}}=\infty$.

In the metric in eq.~\eqref{eq:AdSSmetric}, the surface $\rho=\rho_h$ corresponds to an event horizon and therefore, the coordinate $\rho$ extends between the UV boundary and the horizon, $\rho_h<\rho<\Lambda_{\textrm{UV}}$. Unlike the RS metric in eq.~\eqref{eq:RSmetric}, the AdS-S metric in eq.~\eqref{eq:AdSSmetric} does not have an IR boundary, which is now instead ``hidden'' behind the horizon. Given the previous discussion,
this absence of the IR boundary indicates an absence of confinement. This lets us model the deconfined 4D theory using the dual AdS-S geometry in eq.~\eqref{eq:AdSSmetric}. The temperature of the deconfined plasma is dual to the Hawking temperature corresponding to the horizon, 
\begin{align}\label{hawkingt}
T=\hbar\rho_h/\pi,
\end{align}
where we have momentarily written the factor of $\hbar$ explicitly for later purposes.  Given the absence of the IR boundary, and hence the radion modulus $\phi$, the Goldberger-Wise perturbation does not play any significant role unlike the confined phase and is therefore negelcted to a leading approximation.\footnote{The non-gravitational scalar action $S_\chi$ is not enhanced by a factor of $M_5^3/k^3$ unlike $S_{\textrm{GR}}$, we will ignore the contribution of $S_\chi$ while discussing the deconfined phase.} 

For finite $\Lambda_{\textrm{UV}}$, there exists an elegant exact solution given in \cite{Savonije:2001nd}. From the dual 4D perspective, it corresponds to the 4D deconfined plasma coupled to 4D GR, gravitationally equivalent to a period of FRW radiation dominance. For $\Lambda_{\textrm{UV}}\gg T$, this results in an adiabatic redshifting of $T$ over time. That is to say that even for finite but large $\Lambda_{\text{UV}}$, AdS-S metric in eq.~\eqref{eq:AdSSmetric} represents an approximate solution if we take $\rho_h \rightarrow\rho_h(t)$, determined by the redshifting \eqref{hawkingt} and the Friedmann equations.

In terms of this quasi-static $T$, one can calculate the free energy density of the deconfined phase \cite{Creminelli:2001th},
\begin{eqnarray}\label{eq:fdecon}
F_{\rm deconfined}
=V_0-2\pi^4 M_5^3 T^4,
\end{eqnarray}
 where $V_0$ is a possible constant energy density that will be determined in a moment.\footnote{In the full model the free energy would also include the contribution of the significant number of (relativistic) non-composite elementary fields of the SM, appearing in 5D as zero modes which are not strongly leaning towards the IR. However, in what follows they are largely ``spectators'' which will not alter the leading results. We therefore omit them.}

To compute the critical temperature corresponding to the PT, we also need to compute the free energy of the RS phase. As mentioned earlier, without Goldberger-Wise $S_\chi$ contribution, the radion $\phi$ is a flat direction. Therefore, the leading contribution to the free energy corresponding to the confined phase at low $T$ comes from the perturbation $S_\chi$ in the background of the unperturbed RS metric eq.~\eqref{eq:RSmetric},
\begin{align}\label{eq:GWaction}
S_{\chi} &=  \int d^4x \int_{\Lambda_{\text{IR}}}^{\Lambda_{\text{UV}}} d\rho\sqrt{-g}\left[-\frac{1}{2}(\partial \chi)^2- V_{\chi}(\chi) - \rho\delta(\rho-\Lambda_{\text{UV}}) \kappa  (\chi^2-v^2)^2 + \rho\delta(\rho-\Lambda_{\text{IR}}) \alpha\chi\right]. 
\end{align}
A free Goldberger-Wise field, $V_{\rm GW}(\chi)= \frac{1}{2}m^2 \chi^2$, with boundary terms above such that they satisfy the boundary conditions for the $\chi$ field
\begin{equation}\label{eq:BCs}
\chi|_{\rho=\Lambda_{\text{UV}}}=v~~ \textrm{and}~~ \rho \frac{\partial\chi}{\partial \rho}\Big |_{\rho=\Lambda_{\text{IR}}}=-\alpha,
\end{equation}
is sufficient to stabilize the geometry \cite{Goldberger:1999uk}.\footnote{Note, to get the UV boundary condition we assumed $\kappa\gg 1$.}
The low energy radion effective action (up to two derivative order) is given by eq.~\eqref{eq:RSaction} after promoting $\Lambda_{\text{IR}}$ to the 4D radion field $\phi(x)$. 
In particular, the boundary terms of $S_{\text{GR}}$ in eq.~\eqref{eq:RSaction} gives the kinetic term of the radion, and the Goldberger-Wise action in eq.~\eqref{eq:GWaction} gives effective radion potential $V_{\text{rad}}$ so that the radion action becomes, 
\begin{align}\label{eq:L_radion}
\mathcal S_{\text{radion}} \approx \int d^4x\left(-6M_5^3(\partial\phi)^{ 2}-V_\text{rad}(\phi)\right).
\end{align}
The  effective radion potential is given by \cite{Goldberger:1999uk},
\begin{equation}\label{radionpot1}
V_{\text{rad}}(\phi) = 12M_5^3\lambda \phi^4 \left(1-\frac{1}{1+\epsilon/4} \left( \frac{\phi}{\langle\phi\rangle}\right)^{\epsilon}   \right)+V_1,
\end{equation}
where $\lambda = (\tau_{_{\rm IR}}+12M_5^3-\frac{1}{8}\alpha^2)/(12M_5^3)$ and $\tau_{_{\rm IR}}$ \ is the tension of the IR boundary  which can be de-tuned away from the RS value of $-12M_5^3$. The above potential has a minimum at $\langle \phi \rangle$ which is hierarchically smaller than
$\Lambda_{\text{UV}}$ \cite{Goldberger:1999uk,Chacko:2012sy},
\begin{align}\label{eq:hierarchy_1FP}
\frac{\langle\phi\rangle}{\Lambda_{\text{UV}}} \equiv \frac{\Lambda_{\text{IR}}}{\Lambda_{\text{UV}}} =\left( \frac{12 M_5^3 \lambda}{\alpha v}\right)^{\frac{1}{\epsilon}}.
\end{align}
Here, $V_1$ is a constant energy density that will be determined below. The parameter $\epsilon$ controlling the hierarchy is determined by the mass of the Goldberger-Wise field, 
\begin{align}\label{eq:GWmass}
\epsilon= -2+\sqrt{4+m^2}\underset{m^2\ll 1}{\approx} m^2/4.  
\end{align}
Requiring small back reaction of the Goldberger-Wise field to the AdS-S geometry requires $\lambda,\epsilon<1$. If we think of $\Lambda_{\text{UV}}$ being of the order of highest scales $\sim M_{\text{Pl}}$, and $\Lambda_{\text{IR}}\gtrsim \textrm{TeV}$, so as to solve the Hierarchy Problem, this large hierarchy can emerge from modest parameters with ${12M_5^3\lambda}/(\alpha v)$ and $\epsilon$ being $O(0.1)$. To get the free energy, we note that for $T\ll \phi$, the KK modes are not excited and $\phi(x)$ is the only dynamical light field in the confined phase. Thus the above effective potential also gives the free energy of the confined phase,
\begin{equation}\label{eq:fcon}
F_{\text{confined}}\underset{{T\ll \phi}}{\approx}V_{\text{rad}}(\phi).
\end{equation}
Let us now relate the two constant energy densities appearing in eqs. \eqref{eq:fdecon} and \eqref{eq:fcon}. We require that the two geometries given by eqs. \eqref{eq:RSmetric} and \eqref{eq:AdSSmetric} match when  $\rho_h\rightarrow 0$ and $\phi\rightarrow 0$ since in that case they both describe zero temperature with the IR boundary removed (to $\rho=0$).
Thus the two free energies given in eqs.  \eqref{eq:fcon} and \eqref{eq:fdecon} should also match when $T\rightarrow 0$ and $\phi\rightarrow 0$, thereby making $V_0=V_1$. We can now calculate the temperature at which the two free energies in eqs. \eqref{eq:fdecon} and \eqref{eq:fcon} become equal, namely the critical temperature for the PT, $T_c$:
\begin{align}\label{Tc}
F_{\rm deconfined}(T_c)=F_{\rm confined}(T_c)~\Rightarrow~
\frac{T_c}{\Lambda_{\text{IR}}}= \left(\frac{-6\ep \lambda}{\pi^4(4+\ep) }\right)^{1/4}.
\end{align} 
Thus $T_c$ is parametrically smaller than $\Lambda_{\text{IR}}$ for small $\ep$ and(or) small $\lambda$. This justifies the effective description of the confined phase involving only the radion as well as eq.~(\ref{eq:L_radion}). Furthermore, the above fact indicates that at the temperature $T_c$ we can have a simultaneous existence of both the confined and the deconfined phases, and thus the PT under consideration is first order in nature. It follows from eq. (\ref{Tc}) that $V_0 = 2\pi^4 M_5^3 T_c^4$ for the almost vanishing cosmological constant today i.e. at $\phi=\langle\phi\rangle$.

The semi-classical bounce solution that we will compute in the next section will correspond to quantum tunneling in terms of the gravitational radius $\rho_h$, but in terms of $T$ (to match with CFT expectations), it will correspond to a thermal transition as is clear from the presence of $\hbar$ in eq.~\eqref{hawkingt}.

\section{The structure of the bounce}  \label{sec:bounce}
As discussed in the previous section, the effect of 4D gravity can be approximated by an adiabatic adjustment of the temperature $T$ and the Hubble parameter $H$. Starting at high $T$ and in the AdS-S phase as the universe expands, $T$ eventually drops below $T_c$. As this happens, the RS phase having a smaller free energy, becomes thermodynamically favorable and bubbles of the RS phase can start nucleating. The phase transition completes only after the bubbles start percolating, and this happens when the rate of bubble nucleation per unit volume, $\Gamma$,
gets bigger than $H^4$. For $T \ll T_c$, $H$ asymptotes to a constant given by $H^2\approx \frac{8\pi }{3} G_N V_0 \sim\frac{2\pi^4 M_5^3 T_c^4}{3M_{\text{Pl}}^2}$. Here, $G_N$ and $M_{\text{Pl}}$ are respectively Newton's constant and the reduced Planck scale, $M_{\text{Pl}}=2.4\times 10^{18}$ GeV. To find the temperature at which the phase transition completes, we need to compute the nucleation rate $\Gamma$. Since the 5D theory is weakly coupled, we can use a semiclassical approximation to compute
$\Gamma$ in terms of a Euclidean bounce action $S_{\rm b}$ as,
\begin{equation}\label{Gamma}
\Gamma \sim T^4e^{-S_{\rm b}}.
\end{equation}
Thus for the PT to complete one needs roughly 

\begin{eqnarray}\label{eq:S_estimate}
S_{\rm b} < 4\ln\left(\frac{\mpl}{T_c}\right)\sim 140,
\end{eqnarray}
where we used $T_c\sim\mathcal{O}$(TeV).

\begin{figure}[t!]
 \centering
 \begin{subfigure}[b]{0.8\textwidth}
 \centering
 \includegraphics[width=\textwidth ]{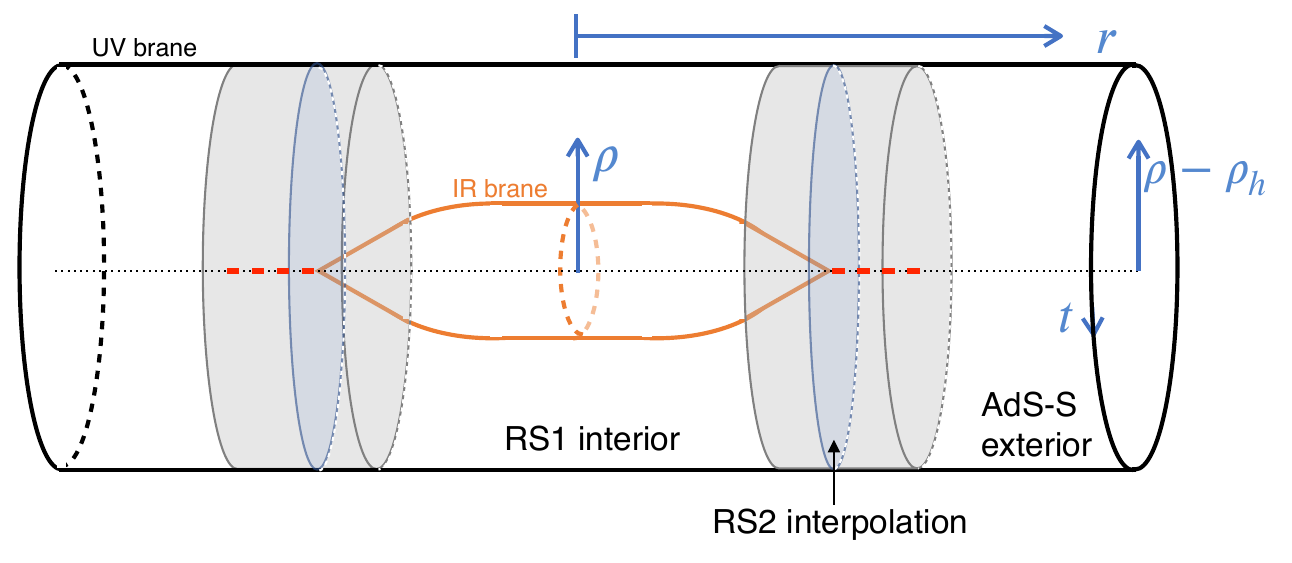}
 \caption{ }
 \label{figure:Ansatz_Rattazzi}
 \end{subfigure}
 
\vspace{4 mm}
\begin{subfigure}[b]{0.8\textwidth}
\centering
\includegraphics[width=\textwidth ]{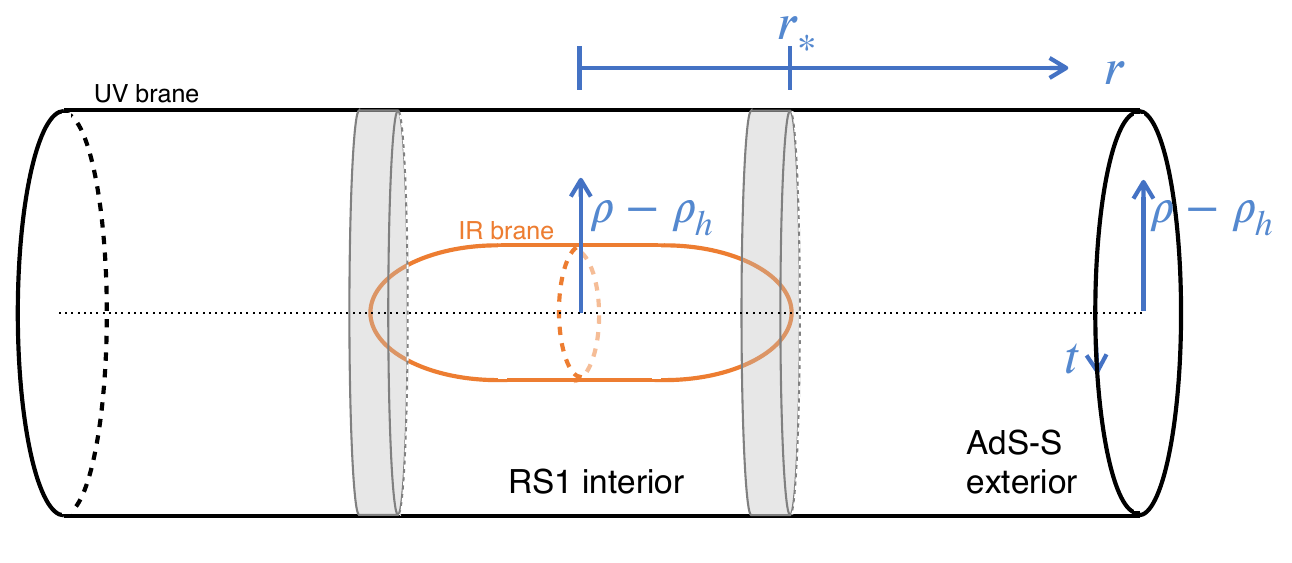}
\caption{ }
 \label{figure:Ansatz_topology}
 \end{subfigure}
 \caption{(a) Topology of the bounce proposed in  ref.~\cite{Creminelli:2001th}. In the gray region  the description of the ansatz gets out of 4D radion EFT as well as 5D EFT control. The orange surface shows the IR brane, and the red dashed line indicates the adjustment of $\rho_h$ between its equilibrium AdS-S value to zero at the RS2 limit. (b) Smooth bounce topology/configuration proposed in this work, describable within 5D EFT. The IR brane (the orange surface) is smoothly capped off at the horizon. In the gray region, the (two-derivative) 4D radion EFT gets out of control, but the bounce can still be described within 5D EFT. }
 \end{figure}

To compute the bounce action, in principle, one has to look for a solution of the Euclidean equation of motion (EoM) derived from the 5D action \eqref{eq:RSaction} that smoothly interpolates between the two above mentioned geometries in eqs.~\eqref{eq:AdSSmetric} and \eqref{eq:RSmetric} (with time compactified on a circle of circumference $1/T$). Authors of ref.~\cite{Creminelli:2001th} attempted to make this trade at the common RS2 limit of both phases  (RS2 is the $\Lambda_{\textrm{IR}}=0$ limit of RS1 and also the limit $T=0$ of AdS-S). They then tried to derive the RS1 phase of the bubble interior by solving the Euclidean EoM within the 4D radion EFT (up to two derivative order). Qualitatively, their results are shown in figure \ref{figure:Ansatz_Rattazzi}. In most of the interior region, the 4D radion EFT bounce is controlled and approximates the true 5D bounce. However, as pointed out in ref.~\cite{Creminelli:2001th}, the central problem with this proposed bounce solution is that it goes out of 4D radion EFT control as one approaches the RS2 juncture (shown in gray in figure~\ref{figure:Ansatz_Rattazzi}) since the IR scale $\phi$ becomes smaller than $T$. Indeed, the lack of smoothness of the brane at the RS2 point takes this bounce configuration outside even 5D EFT control. The interpolation from RS2 to AdS-S is similarly out of the 5D EFT control. 

By contrast, we will consider a \textit{smooth} bounce configuration as illustrated in figure \ref{figure:Ansatz_topology}. As in figure \ref{figure:Ansatz_Rattazzi}, the interior region is described by a bounce solution in 4D radion EFT, but deviates from figure \ref{figure:Ansatz_Rattazzi} for $\phi\sim T$ near the transition regime (shown in gray in figure \ref{figure:Ansatz_topology}), with the brane capped off smoothly so that it is controlled within 5D EFT. 
With this smoothness criterion, which we will elaborate further below, finding the bounce solution is a mathematically well-posed question. However the exact solution is difficult to find in practice.\footnote{Nevertheless, see our remarks in the introduction regarding the possibility of obtaining the actual bounce solution in the thin-wall limit 
 for a 6D example \cite{Aharony:2005bm}, taking advantage of a symmetry between the Euclidean time circle and the sixth dimension. } Instead, we will proceed by making a reasonable  \textit{ansatz} for the 5D geometry of the bounce that satisfies the same smoothness criterion.
Although such an ansatz may not be the true bounce solution, (a) we will argue later that in the thin-wall regime,
\begin{equation}\label{ansatzbound}
S_{\text{b, ansatz}}^\text{thin-wall} > S_{\text{b, true}}^\text{thin-wall},
\end{equation}
so it can provide an upper bound for $S_{\text{b, true}}^\text{thin-wall}$ and hence a lower bound for $\Gamma$; (b) it provides a reasonable estimate for  $\Gamma$ more generally.

 For the smooth ansatz we consider in this paper, the entire configuration can be described as globally AdS-S with the region inside the IR brane ``cut-out''. In particular, the RS1 phase in the far interior is being approximated by the IR brane cutting out a portion of AdS-S rather than a zero temperature AdS. This is a good approximation because in the far interior $\phi\gg T$. The key to smoothness of our ansatz, and hence 5D EFT control, is that the brane is capped off at the horizon $\rho=\rho_h$ as opposed to at $\rho=0$ as in ref.~\cite{Creminelli:2001th}. However, we make no claim that the smooth transition region, shown in gray, is a controlled approximation to the true bounce in that it does not solve the 5D Euclidean EoM. Rather, it is a qualitatively accurate ansatz, which smoothly interpolates controlled approximations in the far interior and the far exterior indicated in white region in figure~\ref{figure:Ansatz_topology}. As pointed out above, in the thin-wall limit this ansatz will provide us a lower bound on the true nucleation rate while outside this limit, the ansatz can be expected to give us a reasonable estimate of this rate.

In more detail, we take our ansatz for the bounce to be described by
\begin{equation}\label{AdS-S}
ds^2 = \left(\rho^2-\frac{\rho_h^4}{\rho^2}\right)dt^2+\frac{d\rho^2}{\rho^2-\frac{\rho_h^4}{\rho^2}}+\rho^2\sum_i dx_i^2,
\end{equation}
with $\rho_h<\phi(r)<\rho<\Lambda_{\rm UV}$ for $r\equiv |\vec{x}|<r_*$ and $\rho_h<\rho<\Lambda_{\rm UV}$ for $r>r_*$. 
Here $\phi(r)$ changes between some release point $\phi_r \gg \rho_h$ at $r=0$ and $\rho_h$ at $r_*$, describing an $r$-dependent IR brane end to the extra dimension in the region $r<r_*$.\footnote{In the thin-wall regime, $\phi_r=\langle\phi\rangle$, but away from the thin-wall regime, $\phi$ generally approaches a  $\phi_r<\langle\phi\rangle$ at $r=0$} We see that in the far exterior this is the AdS-S metric and in the far interior, since  $\phi_r \gg \rho_h$, this approximates RS1 at very low temperatures eq.~\eqref{eq:RSmetric}. This gives an $O(3)$ symmetric ansatz for the bounce, schematically shown in  figure~\ref{figure:Ansatz_topology}. The justification for a time independent $O(3)$ symmetric bounce structure arises as a result of (two derivative) radion dominance of the bounce action and is given in ref.~\cite{Agashe:2019lhy}.  Note that the choice $\phi(r_*)=\rho_h$ implies that the IR brane is pinching off to zero time circumference at $r_*$ so that we have a closed IR brane hypersurface. It remains to choose a specific $\phi(r)$ that this pinching off results in a \textit{smooth} brane embedding seen in figure \ref{figure:Ansatz_topology}. 

We will first show how to compute the action eq.~\eqref{eq:RSaction} for a given $\phi(r)$. We will then discuss the conditions that ensure a smooth ansatz and minimize the action subject to those conditions to obtain the ansatz (i.e. $\phi(r)$) that gives us the strongest upper bound on the bounce action in the thin-wall limit. We now proceed by computing different terms in the action eq.~\eqref{eq:RSaction}. To obtain the extrinsic curvature $K$, we first find the unit normal vector to the surface $\rho = \phi(r)$,
\begin{equation}
n_\phi = \left(\frac{\rho^2}{\rho^4-\rho_h^4+\phi^{\prime 2}}\right)^{1/2} (0,-\phi^\prime,0,0,1),
\end{equation}
where $\phi^\prime\equiv \frac{d\phi}{dr}$.
The induced metric on this surface is given by
\begin{align}
ds_{\text{ind}}^2 = \left(\rho^2-\frac{\rho_h^4}{\rho^2}\right)dt^2 + \left(\rho^2+\frac{\phi^{\prime 2}}{\rho^2-\frac{\rho_h^4}{\rho^2}}\right)dr^2 
+\rho^2(r^2d\theta^2+r^2\sin^2\theta d\varphi^2).
\end{align}
From the above the trace of the extrinsic curvature and the determinant of the induced metric can be calculated as,
\begin{align}
\sqrt{\gamma} = r^2&\sin\theta\phi^2\left(\phi^4-\rho_h^4+\phi^{\prime 2}\right)^{1/2},\\
\sqrt{\gamma} K = r^2&\sin\theta \frac{1}{\phi^4-\rho_h^4+\phi^{\prime 2}}\times \nonumber\\
&\left[2\phi(\rho_h^4-\phi^4)\frac{\phi^\prime}{r}+(6\phi^4-2\rho_h^4)\phi^{\prime 2}-2\phi\frac{\phi^{\prime 3}}{r}
+(\phi^4-\rho_h^4)(4\phi^4-2\rho_h^4-\phi\phi''(r))\right].
\end{align}
Using the above we can evaluate the full action in eq.~\eqref{eq:RSaction} as a function of $\phi(r)$ to get,
	\begin{align}\label{fullaction2}
	S_{\phi} = & \frac{4\pi}{T}\int dr r^2\Bigg[2M_5^3\Big(\frac{2}{\phi^4-\rho_h^4+\phi^{\prime 2}}
	\Big[2\phi(\rho_h^4-\phi^4)\frac{\phi^\prime}{r}+(6\phi^4-2\rho_h^4)\phi^{\prime 2}-2\phi\frac{\phi^{\prime 3}}{r} \nonumber\\
	&+(\phi^4-\rho_h^4)(4\phi^4-2\rho_h^4-\phi\phi'')\Big] +\rho_h^4-2\phi^4-6\phi^2\left(\phi^4-\rho_h^4+\phi^{\prime 2}\right)^{1/2}\Big) +   V_{\text{rad}}(\phi)\Bigg],
	\end{align}
where $V_{\text{rad}}(\phi)$ is the contribution coming from $ S_\chi$. For $\rho>\phi\gg \rho_h$ (the white interior region in figure~\ref{figure:Ansatz_topology} outside the orange surface), the AdS-S geometry, the low temperature RS geometry, and the zero temperature RS geometry are all approximately the same and $V_{\text{rad}}(\phi)$ becomes the standard zero temperature radion potential given in eq. \eqref{radionpot1}. In this regime and for $\phi^\prime\ll \phi^2$, the effective action \eqref{fullaction2} reduces to the standard two derivative radion action in the Euclidean time independent $O(3)$ symmetric regime, 
\begin{align}\label{eq:action_radion}
S_\phi \approx\frac{4\pi}{T}\int dr r^2\Big[6M_5^3\phi^{\prime 2}+V_\text{rad}(\phi)\Big].
\end{align}
 In this $\phi \gg \rho_h$ region, the solution to the EoM derived from the above 
action is a controlled approximation to the true bounce and gives the parameterically dominant contribution to the bounce action, as anticipated in refs.~\cite{Creminelli:2001th,Agashe:2019lhy} and fully justified in section~\ref{sec:thinwall}.  However, the solution itself  takes us out of radion EFT into the region $\phi\sim\rho_h$ (shown in gray in figure \ref{figure:Ansatz_topology}).
For $\phi\sim\rho_h$, although it is not straightforward to calculate the contribution of $S_{\chi}$ in detail, its contribution to $S_\phi$ is suppressed by the  parameter $\lambda$ (see eq. \eqref{radionpot1}), whereas the rest of the terms in $S_\phi$ are \textit{un}suppressed, as can be seen using eq. \eqref{fullaction2}. Thus for $\phi\sim \rho_h$, $V_{\text{rad}}(\phi)$ will not play a significant role in determining the bounce and we will keep using the same $V_{\text{rad}}(\phi)$ given in eq. \eqref{radionpot1} even in this region for convenience. The contribution of $S_{\chi}$ to terms involving $\phi^\prime$ is also suppressed by $\lambda$ compared to terms arising from 5D gravitational action $S_{\textrm{GR}}$ and thus has been neglected in eq. \eqref{fullaction2}.

\begin{figure}
	\centering
	\includegraphics[width=0.75\linewidth]{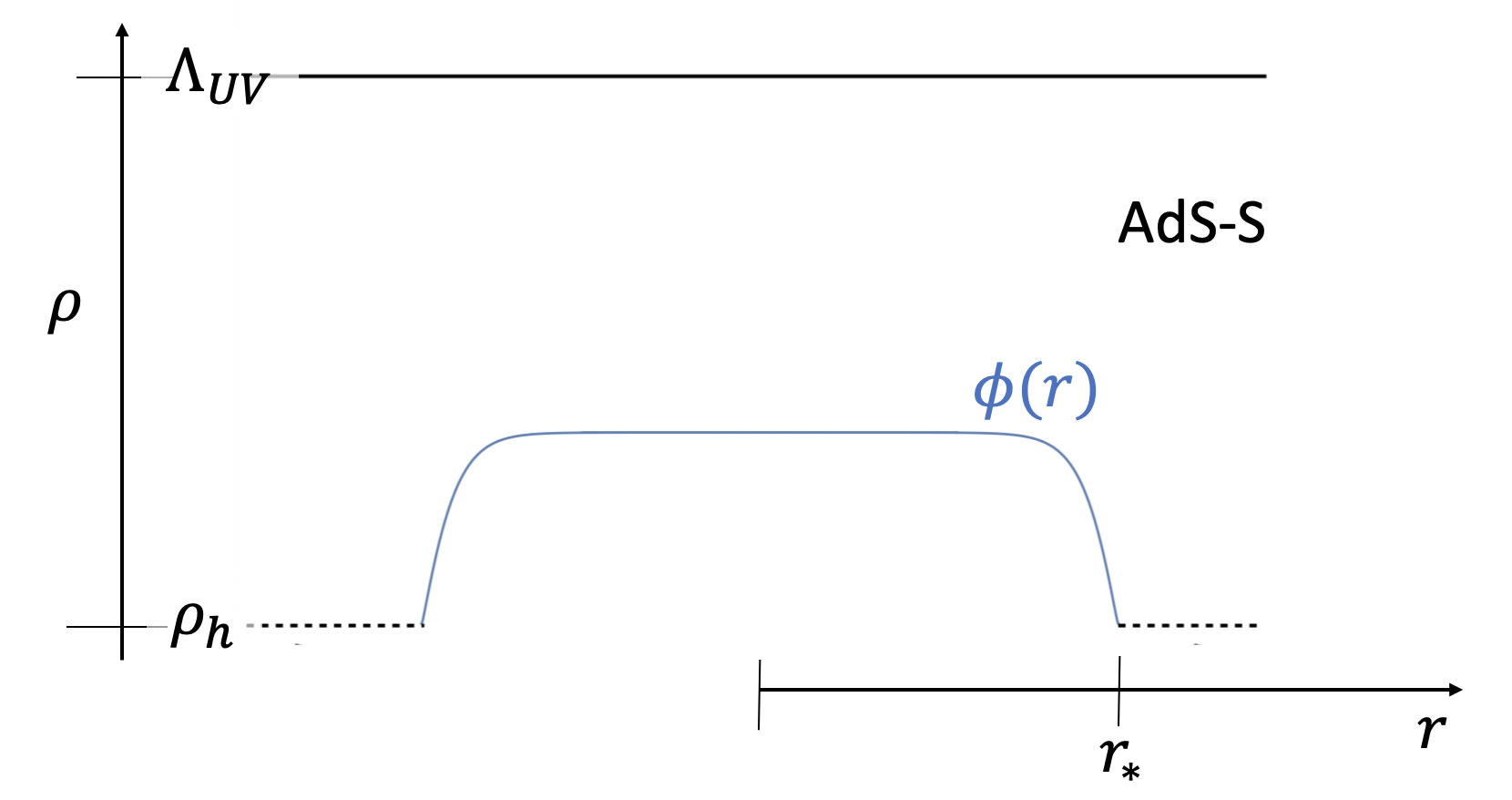}
	\caption{ An example of the profile of the bounce ansatz, obtained by solving the EoM resuling from the action of eq.~\eqref{fullaction2}. The black dashed lines represents the black-brane horizon and the blue curve shows the IR-brane profile specified by $\phi(r)$.}
	\label{figure:bubble-profile}
\end{figure}

To find the optimal form of $\phi(r)$ for our anstaz (eq.~\ref{AdS-S} and discussion below it, figure~\ref{figure:Ansatz_topology}) in the thin-wall limit we will choose $\phi(r)$ to minimize the bounce ansatz action eq. \eqref{fullaction2}  by solving the ``EoM'' that follows from it. We show an example of such a solution in figure \ref{figure:bubble-profile}.
For this solution, the 5D geometry is smooth everywhere except in the potentially problematic region where the IR boundary merges into the AdS-S horizon ($r\approx r_*$ region in figures \ref{figure:Ansatz_topology} and \ref{figure:bubble-profile}). To see whether this merging is smooth, we evaluate the induced metric in the near-horizon region by writing $\phi(r) = \rho_h+\delta\phi(r)$. Assuming $\rho_h^3\delta\phi\ll \phi'^2$, we get
\begin{align}
ds_{\text{ind}}^2\supset 4\rho_h\delta\phi dt^2+\frac{\phi'^2}{4\rho_h\delta\phi}dr^2 
=4\rho_h^2y^2dt^2+dy^2,
\end{align}
where we have made the change of variable/coordinate $y = \sqrt{\delta\phi/\rho_h}$. Note that this is the same as the metric of a flat space with the correct time periodicity. This piece of flat space is embedded in AdS-Schwarzchild, with (near horizon) metric
\beq
ds^2=4\rho_h^2y^2dt^2+dy^2+\rho_h^2 \left( dr^2 + r^2 \left( d\theta^2 + \sin^2 \theta d \varphi^2 \right) \right) ,
\eeq
at a fixed $r$. This ensures a smooth brane, smoothly embedded in AdS-S, with two coordinates/directions $t$ and $r$ acting as spectators, analogous to embedding of a 2D sphere in 3D flat space.

For concreteness, we will choose the boundary condition
\begin{eqnarray}\label{eq:boundary_condition}
\phi'\vert_{\phi=\rho_h}=\rho_h^2
\end{eqnarray}
for our ansatz which respects the above condition $\rho_h^3\delta\phi\ll \phi'^2$ near the horizon, ensuring a smooth brane embedding. This condition, along with the usual smoothness condition at the bubble center, $\phi'\vert_{r=0}=0$, fixes our ansatz.\footnote{ We note that although the action eq. \eqref{fullaction2} involves $\phi''$, the EoM obtained from it is still a second order differential equation for $\phi(r)$ and so (only) two boundary conditions are needed to fix its solution.}

\section{Phase transition in thin-wall regime}
\label{sec:thinwall}
Having discussed the effective action \eqref{fullaction2} for $\phi$  and the boundary conditions needed to fix the solution, we can now calculate the bounce action based on our ansatz. Although one can proceed numerically in general,
in the thin-wall regime ($T\approx T_c$) it is possible to obtain an analytical expression for the bounce action.  

As mentioned before, we focus on the bounce action with $O(3)$ symmetry. The $O(3)$ symmetric bounce action can be rewritten quite generally in the thin-wall regime as \cite{Coleman:1977py,Linde:1981zj},
\begin{equation}\label{eq:S_3_S_1}
S_{\rm b}=\frac{S_3}{T} = \frac{16 \pi}{3}\frac{S_1^3}{(\Delta F)^2 T},
\end{equation}
where  $\Delta F$ is the difference of the free energy in two phases. $S_1$ is the surface tension of the bubble wall, evaluated in the degenerate limit $\Delta F=0$. Therefore, any configuration of $S_1$ is always bounded from below. Finding the true solution of $S_1$ is basically minimizing $S_1$. Any other ansatz, which is not the solution, should satisfy
$S_{1,\text{ansatz}}>S_{1,\text{true}}$. Combining with eq. \eqref{eq:S_3_S_1}, this leads to $S_{\rm b,\text{ansatz}}>S_{\rm b,\text{true}}$. However, a random $S_{\rm b,\text{ansatz}}$ might
involve singularities which take us outside 5D EFT control.
In this regard, our smooth ansatz (figure~\ref{figure:Ansatz_topology}) provides a controlled upper bound on the true bounce action in the thin-wall limit.

For our ansatz, the surface tension of the bubble $S_1$ can be obtained from eq. \eqref{fullaction2} as,
	\begin{align}\label{s1action}
	S_1 &= \int dr \Bigg[2M_5^3\Bigg(\frac{2	\Big[(6\phi^4-2\rho_h^4)\phi^{\prime 2}
	+(\phi^4-\rho_h^4)(4\phi^4-2\rho_h^4-\phi\phi'')\Big]}{\phi^4-\rho_h^4+\phi^{\prime 2}}
 \nonumber\\
	&~~+\rho_h^4-2\phi^4-6\phi^2\left(\phi^4-\rho_h^4+\phi^{\prime 2}\right)^{1/2}\Bigg) +   V_{\text{rad}}(\phi)\Bigg].
\end{align}
To clearly show the parametric dependence of $S_1$, we divide it into two parts: $S_1=S_1^{\rm radion}+S_1^{\rm transition}$, where $S_1^{\rm radion}$ is the part of bounce action deep inside the bubble (see the white interior of figure~\ref{figure:Ansatz_topology}), where 4D radion EFT (eq.~\eqref{eq:action_radion}) is valid, while $S_1^{\rm transition}$ denotes the contribution in the transition region $\phi\sim \rho_h$, where the 5D EFT is needed (the gray region of figure~\ref{figure:Ansatz_topology}). As shown around eq. \eqref{eq:action_radion}, in the interior of the bubble, where $\phi\gg \rho_h$ and $\phi^2\gg\phi'$, $S_1^{\rm radion}$ reduces to the one dimensional radion action 
\begin{eqnarray}\label{eq:S_1_radion_dominance}
    S_1^{\rm radion}\approx\int_0^{\lesssim r_*} dr \Big[6M_5^3\phi^{\prime 2}+V_\text{rad}(\phi)\Big]&
    \approx&\sqrt{24 M_5^3}\int_{\gtrsim\rho_h }^{\langle\phi\rangle}d\phi \sqrt{V_{\text{rad}}(\phi)}\nonumber\\
    &\approx& 0.9\,(16 \pi^2 M_5^3) \left(\frac{1}{\epsilon \lambda}\right)^{1/4} T_c^3~~~(\epsilon\lambda\ll 1).
\end{eqnarray}
where the equality in the first line follows from the EoM. One of the integration limits is fixed because the bounce action starts at the minimum of the potential ($\phi=\langle\phi\rangle$). As $\phi$ approaches $\rho_h$, the approximation in eq. \eqref{eq:S_1_radion_dominance} breaks down and therefore we stop the integration at $\phi\gtrsim \rho_h$. In the second line of eq. \eqref{eq:S_1_radion_dominance}, we used eq. \eqref{Tc} to rewrite $\langle\phi\rangle$ in terms of $T_c$. It is clear from eq. \eqref{eq:S_1_radion_dominance} that $S_1^{\rm radion}$ is enhanced for small $\epsilon$ and/or $\lambda$.

The expression for $S_1^{\rm transition}$ is hard to obtain analytically but one can easily get its parametric dependence. Given $\phi\sim \rho_h$ and $\phi'\sim \rho_h^2$ in the transition region, all terms in $S_1^{\rm transition}$ are of order $M^3_5 \rho_h^3$ (see eq. \eqref{s1action}). Therefore, $S_1^{\rm transition}\sim M_5^3 T_c^3$ and it is parametrically smaller than $S_1^{\rm radion}$ for small $\epsilon$ and/or $\lambda$. Now we can conclude that, for small $\epsilon$ and/or $\lambda$,  $S_1\approx S_1^{\rm radion}$, which means the bounce action is dominated by the contribution from standard radion EFT (up to two derivative order).
Plugging $S_1^{\rm radion}$ in to eq. \eqref{eq:S_3_S_1}, $S_{\rm b}$ is therefore given as
\begin{eqnarray}\label{eq:thinbounce}
    S_{\rm b}\approx8\, (16 \pi^2 M_5^3) \left(\frac{1}{\epsilon \lambda}\right)^{3/4} \frac{T_c/T}{\left((1-\left(T/T_c\right)^4\right)^2}~~~~~(\epsilon\lambda\ll 1).
\end{eqnarray}
The radion EFT dominance is also justified based on some benchmark points in table \ref{tab:S_1_comparison}.

\begin{table}[h]
    \centering
    \begin{tabular}{c|c|c|c}
    \hline
    \hline
      Model parameters   & $S_1/(16\pi^2 M_5^3T_c^3)$ & $S_1^{\rm radion}/(16\pi^2 M_5^3T_c^3)$ & $S_{\rm b}/(16\pi^2 M_5^3)$ \\
      \hline
      \hline
      $\epsilon=1/2,\lambda=1/2$ & 1.2 & 0.5 & 90 \label{line1} \\
       \hline
        $\epsilon=1/25,\lambda=1/2$ & 2.1 & 1.3 & 486 \\
         \hline
       $\epsilon=1/25,\lambda=1/25$ & 4.1 & 3.3 & $3.7\times 10^3$ \\
       \hline
        $\epsilon=1/100,\lambda=1/100$ & 8.7 & 7.8 & $3.4\times 10^4$ \\
        \hline
        \hline
    \end{tabular}
    \caption{Comparison of numerical results of $S_1$ (eq.~\eqref{s1action}) and $S_1^{\rm radion}$ (eq.~\eqref{eq:S_1_radion_dominance}) for different model parameters $\epsilon\,,\,\lambda$. To get the concrete number for $S_1^{\rm radion}$, we set the lower integration limit to $\rho_h$ in the first line of eq.~\eqref{eq:S_1_radion_dominance}. We also show the full bounce action $S_{\rm b}$ (eq.~\eqref{eq:S_3_S_1})  in the thin-wall limit in terms of  $S_1$ at $(T/T_c)^4=1/2$. }
    \label{tab:S_1_comparison}
\end{table}
We see that, while for very small $\epsilon$ and $\lambda$, radion dominance gives a quantitatively good approximation, for small but not very small $\epsilon$ and $\lambda$, the approximation is poor. In these cases of interest, our ansatz in the full 5D theory is key to providing a rigorous bound on the true bounce action.

In the above, we have described the general considerations of PT dynamics in the thin-wall limit. However, when we apply them to the real world, we find that the thin-wall limit is incompatible with the observed Planck-Weak hierarchy. Concretely, with $\lambda \lesssim 1$ to ensure  a controlled back-reaction, and $\epsilon = 1/25$ to obtain the large Planck-Weak hierarchy, we see that the PT does not complete near $T_c$ even for $16\pi^2M^3_5=1$ as can be seen using eq.~\eqref{eq:S_estimate}. However, for theoretical control of the 5D EFT we need $16\pi^2M^3_5>1$.\footnote{This is just a requirement that the quantum gravity loops or $\sim 1/(16\pi^2M^3_5)$ corrections are small at the AdS curvature scale.} 

However, non-minimal models can improve the compatibility with the thin-wall limit. To allow the PT to happen for larger values of $M_5^3$, we need larger values of $\epsilon$ (see the first line of table~\ref{tab:S_1_comparison}) while still generating the correct value of the Planck-Weak hierarchy. We will introduce a scenario which achieves these goals in the next section.

\section{5D realization of a two-FP model \label{sec:twoFP}}
In ref.~\cite{Agashe:2019lhy}, we proposed a scenario from the dual (near-)CFT perspective with distinct UV and IR fixed points, which can simultaneously achieve a large Planck-Weak hierarchy and also have a larger $\epsilon$ controlling the PT, such that it can complete promptly in a theoretically controlled parameter regime. Let us briefly review that argument before proceeding with an explicit 5D model realizing (the dual of) this scenario.

\subsection{4D near-CFT description}
The form  of the potential in eq.~\eqref{radionpot1} can be obtained just by using the approximate conformality of the dual 4D theory. A massive Goldberger-Wise field is dual to a marginal deformation $\Delta\mathcal{L}_{\rm CFT} = \kappa\mathcal{O}$, where $\mathcal{O}$ has scaling dimension $4+\epsilon$ given by eq.~\eqref{eq:GWmass} and $\kappa$ is the running coupling. Here we consider $\epsilon>0$, which means the deformation is sightly irrelevant. In the far-IR this deformation becomes negligible and the theory flows back to the exact CFT FP. This gives rise to a potential for the dilaton $\phi$ \footnote{We use $\phi$ to denote  both radion and dilaton. Since dilaton is dual to the radion, this notation should not cause any confusion.} \cite{ArkaniHamed:2000ds,Rattazzi:2000hs,Goldberger:1999uk},
of the form
\begin{align}\label{eq:dilation_potential}
V_{\rm dilaton}(\phi) \approx \lambda_1\phi^4 + 
\lambda_2\kappa_{\rm UV}\phi^4\left(\frac{\phi}{\Lambda_{\rm UV}}\right)^{\epsilon},
\end{align}
where $\lambda_{1,2}$ are dimensionless parameters which are typically of order unity. $\kappa_{\rm UV}$ is the coupling $\kappa$ evaluated at $\Lambda_{\rm UV}$.
The first term on the right-hand side (RHS) is conformally invariant, but the second term reflects the small $\epsilon-$breaking of conformal invariance by the deformation $\kappa$.

On the other hand, in the far-UV, the coupling $\kappa$ becomes significant. We will consider the possibility that it asymptotes in the UV to a non-trivial FP at $\kappa_*$. We can summarize the near FP behaviors in RG language,
\begin{align}\label{eq:RG_running_kappa}
\frac{d\kappa}{d\ln\mu} \approx \begin{cases} 
   \epsilon\kappa & \text{for small $\kappa$}\\
   \epsilon^\prime (\kappa_*-\kappa) & \text{for $\kappa$ near $\kappa_*$ },
  \end{cases}
\end{align}
where $\mu$ is the RG scale and we consider $\epsilon'>0$. Crudely, there is an intermediate scale $\mu_{\rm match}$
where RG evolution transits from the basin of the UV FP (lower line of eq.~\eqref{eq:RG_running_kappa}) to that of the IR FP (upper line of eq.~\eqref{eq:RG_running_kappa}).

Starting with a UV value $\kappa_{\rm UV}\sim \kappa_*$, the coupling slowly evolves under the influence of the UV FP at $\kappa_*$ until the vicinity of a ``matching scale'' $\mu_{\rm match}\sim \Lambda_{\rm UV}\left(\frac{\kappa_*-\kappa_{\rm UV}}{\kappa_*-\kappa_{\rm match}}\right)^{1/\epsilon^{\prime}}$ where $\kappa_{\rm match}\equiv\kappa(\mu_{\rm match})\sim \kappa_*$. The contribution to dilaton potential from this region will be similar to that in eq.~\eqref{eq:dilation_potential} but with the replacement of $\epsilon \to -\epsilon'$. Below the scale $\mu_{\rm match}$, the coupling evolves under the influence of the IR FP at $\kappa=0$. Correspondingly, for $\phi<\mu_{\rm match}$, the dilaton potential is given by eq.~\eqref{eq:dilation_potential} with the replacements: $\Lambda_{\rm UV}\rightarrow\mu_{\rm match}$ and $\kappa_{\rm UV}\rightarrow\kappa_{\rm match}$. Combining the contributions from these two regions, the dilaton potential in this model can be shown as
\bea\label{eq:dilaton_pot_2FP}
V_{\rm dilaton} (\phi) \approx
\begin{cases}
\lambda_1^\prime\phi^4 + 
\lambda_2^\prime\left(\kappa_*-\kappa_{\rm UV}\right)\phi^4\left(\frac{\phi}{\Lambda_{\rm UV}}\right)^{-\epsilon^\prime} & \phi\gg \mu_{\rm match} \\
\lambda_1\phi^4 + 
\lambda_2\kappa_{\rm match}\phi^4\left(\frac{\phi}{\mu_{\rm match}}\right)^{\epsilon} & \phi\ll \mu_{\rm match}, \\
\end{cases}
\eea 
with $\mathcal{O}(1)$ uncertainties for $\phi\sim \mu_{\rm match}$.

Now we are ready to illustrate why the two-FP scenario illustrated above achieves our goal stated at the beginning of this section. By choosing $\epsilon^\prime\sim 1/25$, we can ensure a sufficiently slow running such that $\mu_{\rm match}$ is hierarchically smaller than $\Lambda_{\rm UV}$.
It is straightforwardly checked that for  $\lambda_1<0,(4+\epsilon)\lambda_{2}\kappa_{\rm match}>-4\lambda_1$ and for either set of conditions:
(i) $\lambda_1',\lambda_2'(\kappa_*-\kappa_{\rm UV})>0$ or (ii) $\lambda_1'>0$ and $-(4-\epsilon')\lambda_2'(\kappa_*-\kappa_{\rm UV})>4\lambda_1'$
, the minimum of the potential $\langle\phi\rangle$ lies in the IR region ($\langle\phi\rangle <\mu_{\rm match}$) with 
\begin{align}\label{eq:phi_VEV_two_FP}
\langle\phi\rangle\sim& \mu_{\rm match} \left(-\frac{\lambda_{1}}{\lambda_{2}\kappa_{\rm match}}\right)^{1/\epsilon}\nonumber\\ \sim& \Lambda_{\rm UV}\left(\frac{\kappa_*-\kappa_{\rm UV}}{\kappa_*-\kappa_{\rm match}}\right)^{1/\epsilon^{\prime}}\left(-\frac{\lambda_{1}}{\lambda_{2}\kappa_{\rm match}}\right)^{1/\epsilon}. 
\end{align}

The above choice of $\epsilon^\prime \ll 1$ ensures a large Planck-Weak hierarchy given $(\kappa_*-\kappa_{\rm UV}) \lesssim (\kappa_*-\kappa_{\rm match})$. The final factor in the second line of eq.~\eqref{eq:phi_VEV_two_FP} will only be modestly small for $\epsilon\lesssim 1$. It is this larger IR exponent $\epsilon$ that will now control the PT dynamics and make them relatively prompt. This is because, as can be seen from
figure~\ref{figure:bubble-profile} and the discussion in section~\ref{sec:bounce}, the PT dynamics is governed by the region $\phi\leq\langle\phi\rangle<\mu_{\rm match}$ and hence entirely by the IR FP basin.

In the rest of this section, we will realize the above scenario in a simple explicit 5D model utilizing an \textit{interacting} Goldberger-Wise stabilizing field, and show that it indeed leads to a relatively prompt PT consistent with a large Planck-Weak hierarchy. This will enable us to obtain explicit expressions of various 4D parameters such as $\epsilon,\epsilon^\prime$ etc. described above, in terms of fundamental 5D parameters, utilizing which we will calculate the rate of the PT.

\subsection{5D model}\label{subsec:twoFP_5D}
We consider the following Goldberger-Wise potential,
\beq\label{eq:V_GW_general}
V_{\chi}= \frac{1}{2} m^{\prime 2} \chi^2 + \frac{1}{6} \eta \, \chi^3 +\frac{1}{24} g \, \chi^4,
\eeq
where $m'$ is the mass of the Goldberger-Wise scalar and $\eta$, $g$ are two coupling constants. The EoM for the extra-dimensional profile of the Goldberger-Wise field in the RS metric of eq.~\eqref{eq:RSmetric} is given by
\begin{align}
\rho^5\partial_\rho^2\chi+5\rho^4\partial_\rho\chi-\rho^3V_\chi'=0.
\end{align}
For later purposes it is convenient to do a coordinate transformation, $\rho=\Lambda_{\rm UV}e^{-\sigma}$ with $0<\sigma<L\equiv \ln (\Lambda_{\rm UV}/\Lambda_{\rm IR})$, following which the EoM reads as,
\begin{align}
    \partial_\sigma^2\chi-4\partial_\sigma\chi-V_\chi'=0.
\end{align}
The above EoM indicates that there exist FP solutions in the extra-dimensional evolution of the Goldberger-Wise profile. They appear where $\chi=\text{constant}:V'_\chi(\chi)=0$ and are located at non-negative values
\begin{align}
    \chi=0~\textrm{and}~\chi=\chi_*=\frac{3}{g}\left(-\frac{\eta}{2}+\sqrt{\frac{\eta^2}{4}-\frac{2gm^{\prime 2}}{3}}\right),
\end{align}
for $m^{\prime 2},\eta<0$ and $g>0$. 

Obtaining the analytical expression for the full radion action from the general Goldberger-Wise potential in eq.~\eqref{eq:V_GW_general} is challenging. However, we can obtain the relevant qualitative insights about the behavior of the Goldberger-Wise profile under eq.~\eqref{eq:V_GW_general} by using the two-FP intuition described in the previous subsection.

In the proximity of the FPs at $\chi=0$ and $\chi=\chi_*$, the field profile evolves quite slowly, whereas as it evolves rapidly near the transition from the vicinity of one FP to the other. Hence, as a simple approximation, we can split the evolution of $\chi$ into two regimes. In the first regime i.e. for field values between $0\leq\chi<\chi_{\rm m}$, we consider the evolution to be governed by the FP at $\chi=0$, and hence with a potential $\frac{1}{2}m^{\prime 2}\chi^2$. For the other segment i.e. for field values between $\chi_{\rm m}<\chi\leq\chi_*$, the evolution is governed by the FP at $\chi=\chi_*$ and hence with a potential $\frac{1}{2}m^{ 2}(\chi-\chi_*)^2$ where $m^{ 2} \equiv V_\chi^{''}(\chi_*)=-2m^{\prime 2}-\eta \chi_*/2$. 
We choose to match two regimes at a field value $\chi_{\rm m}=(-\eta+\sqrt{\eta^2-2gm^{\prime 2}})/g$, which is the inflection point of $V(\chi)$.
To summarize, we consider the simplified Goldberger-Wise potential $\tilde{V}_{\chi}$, \bea\label{eq:GWsimple}
\tilde{V}_{\chi}(\chi)\approx
\begin{cases}
      \frac{1}{2} m^{\prime 2} \chi^2  & \text{for} \; 0\leq\chi<\chi_{\rm m} \\
      \frac{1}{2}m^{ 2} (\chi-\chi_*)^2 - C & \text{for} \; \chi_{\rm m}<\chi\leq\chi_*.
\end{cases} 
\eea
In the above, $C=\frac{1}{2}m^{2} (\chi_{\rm m}-\chi_*)^2-\frac{1}{2} m^{\prime 2} \chi_{\rm m}^2 $ is a constant that ensures the continuity of $\tilde{V}_\chi$ at $\chi_{\rm m}$.

\begin{figure}[h]
\centering
\includegraphics[width=0.75\linewidth]{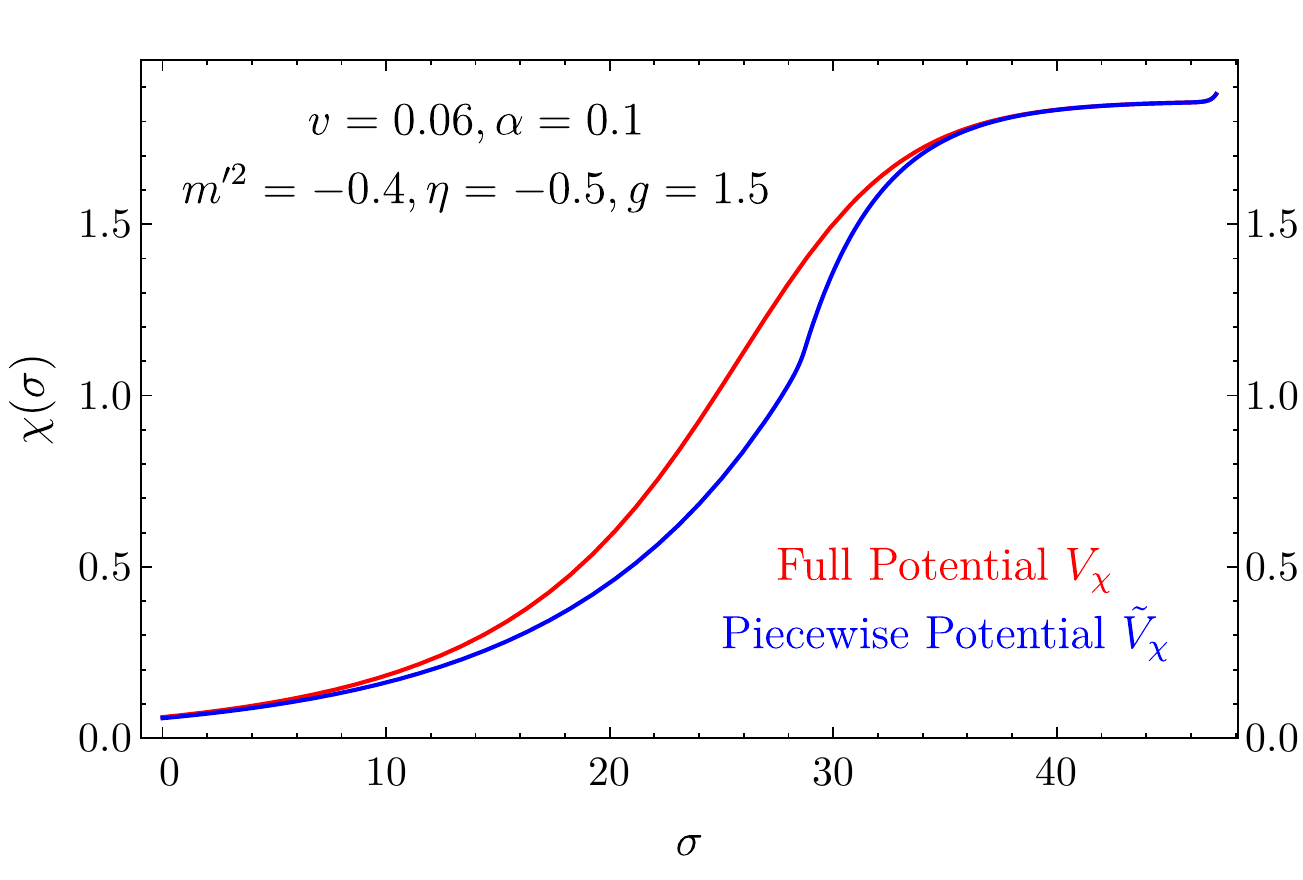}
\caption{The profile of Goldberger-Wise scalar $\chi(\sigma)$ from two different potentials: full potential eq.~\eqref{eq:V_GW_general} and simplified potential eq.~\eqref{eq:GWsimple} . The parameters we choose are $m^{\prime 2}=-0.4$, $\eta=-0.5$, $g=1.5$ and $v$=0.06, $\alpha=0.1$ and $L=15\pi$.}
\label{figure:profile}
\end{figure}

A comparison of the numerical solution obtained for the full potential in eq.~\eqref{eq:V_GW_general} with the simplified potential in eq.~\eqref{eq:GWsimple} is shown in figure~\ref{figure:profile}. Here we choose the same boundary conditions as in eq.~(\ref{eq:BCs}) with $0<v<\chi_{\rm m}$, $\alpha > v$ and fix $L=15\pi$. Figure~\ref{figure:profile} clearly shows the asymptotic behaviour of $\chi(\sigma)$ near 
 $\chi=0$  and $\chi_{*}$, which act as the UV and IR fixed points with respect to the flow in the
extra dimension that is dual to the RG flow in the CFT perspective. As can be seen from figure~\ref{figure:profile}, the simplified potential gives a good approximation to the full potential, and correspondingly we can trust the analytical calculation of the radion potential using eq.~\eqref{eq:GWsimple}.

Given the quadratic piecewise Goldberger-Wise scalar potential in eq.~\eqref{eq:GWsimple},  the radion potential can be calculated analytically (see details in appendix \ref{app:radion_pot_5D}), which can be summarized as 
\bea \label{eq:radpot2FP}
V_{\rm rad} (\phi) \approx
\begin{cases}
 \tau \phi^4  -  \alpha v \phi^{4-\epsilon'} & \phi> \phi_{\rm m} \\
(\tau-\alpha \chi_*) \phi^4 +  
\frac{\alpha\chi_{\rm m}}{2}  \left( \frac{\chi_{\rm m}(1-\frac{\ep}{8})}{v} \right)^{\epsilon/\epsilon'} \phi^{4+\epsilon} & \phi \ll \phi_{\rm m} \\
\end{cases},
\eea
  The above form of the radion potential is the 5D radion realization of the CFT dilaton potential structure in eq.~\eqref{eq:dilaton_pot_2FP}. In eq.~\eqref{eq:radpot2FP}, $\tau = \delta \tau_{\rm IR} - \frac{1}{8}\alpha^2$ is determined in terms of detuning on the IR boundary $\delta \tau_{_{\rm IR}}\equiv \tau_{\rm IR}+12M_5^3$, $\epsilon'=
2-\sqrt{4+m'^2}$ and $\epsilon$ is defined in eq.~\eqref{eq:GWmass}.
Finally, $\phi_{\rm m}\equiv e^{-L_{\rm m}}$ where $L_{\rm m}$ is the smallest size of the extra dimension where the Goldberger-Wise profile reaches the value $\chi_{\rm m}$ on the IR boundary. For $L<L_{\rm m}$, $\chi(\sigma)$ never grows to reach $\chi_{\rm m}$ in the extra dimension, whereas for $L>L_{\rm m}$, $\chi(\sigma)$ becomes bigger than $\chi_{\rm m}$ and goes to the vicinity of the IR FP near the IR boundary.

We can choose the parameters such that the radion potential above has only one minimum for $\phi<\phi_{\rm m}$ at 
\beq\label{eq:radmin}
\langle\phi\rangle\sim \left( \frac{v}{\chi_{\rm m}(1-\ep/8)} \right)^{1/\epsilon'} \left(\frac{- 2\tau'}{ \alpha \chi_{\rm m}}\right)^{1/\epsilon},
\eeq
where $\tau'\equiv \tau- \alpha \chi_*<0$. 
This expression of the hierarchy is to be compared with eq.~\eqref{eq:hierarchy_1FP} in the single-FP scenario. The large Planck-Weak hierarchy can be obtained from the first factor of the RHS of eq.~\eqref{eq:radmin} with a small $\epsilon^\prime \sim 1/25$ and a modest ratio of $v/\chi_*$,\footnote{As we explain in appendix \ref{app:radion_pot_5D}, the radion potential in eq.~\eqref{eq:radpot2FP} has been obtained with the approximation that the second factor of the RHS of eq.~\eqref{eq:radmin} is $\lesssim 0.1$.} while now allowing $\epsilon$ to be considerably larger. As shown in section~\ref{sec:bounce},  it is the region $\phi\lesssim\langle\phi\rangle<\phi_{\rm m}$ that is relevant for PT dynamics and is controlled by $\epsilon$.
Consequently, our computation of the bounce action obtained in the previous sections using eq.~\eqref{radionpot1} can be applied, but now with   
only modestly small $\epsilon\lesssim1$, thereby achieving the goal stated at the end of section \ref{sec:thinwall}.
Correspondingly for $\epsilon > \epsilon^\prime$, the bounce action in the thin-wall regime becomes parametrically smaller, as suggested by eq.~\eqref{eq:thinbounce}, and this allows for the PT to complete for parametrically larger $N$.

As discussed earlier, the bounce in the thin-wall regime encompasses $\phi\leq\langle\phi\rangle<\phi_{\rm m}$.
Thus we can directly use the second line of eq.~\eqref{eq:radpot2FP} to calculate the bounce. Given the exact similarity between this potential and the one in eq.~\eqref{radionpot1}, used to calculate bounce action in section~\ref{sec:thinwall}, we can directly re-use the results given in table~\ref{tab:S_1_comparison}, even though such results were obtained in a single-FP scenario.

Having worked out these general features of the thin-wall regime in our two-FP scenario, with the rigorous bounds following from our ansatz, we now apply them to the case of a realistic PT consistent with the Planck-Weak hierarchy. For a
benchmark set of parameters $\epsilon = 0.5,{\lambda}=0.5$, as shown in table~\ref{tab:S_1_comparison},
the PT can complete promptly for $16\pi^2M_5^3\gtrsim1$ in the thin-wall regime, (marginally) under theoretical control.
This was completely impossible in the original single-FP scenario mentioned in section~\ref{sec:thinwall}.

For the regions of parameter space that the PT does not complete near $T_c$, the universe keeps cooling down to temperatures where the thin-wall approximation is no longer valid. In the next section we will consider such transition temperatures outside the thin-wall regime, with even better semi-classical control.

\section{Phase transition outside thin-wall regime}\label{sec:PT_not_thin_wall}

\begin{figure}[h]
\centering
\includegraphics[width=0.75\linewidth]{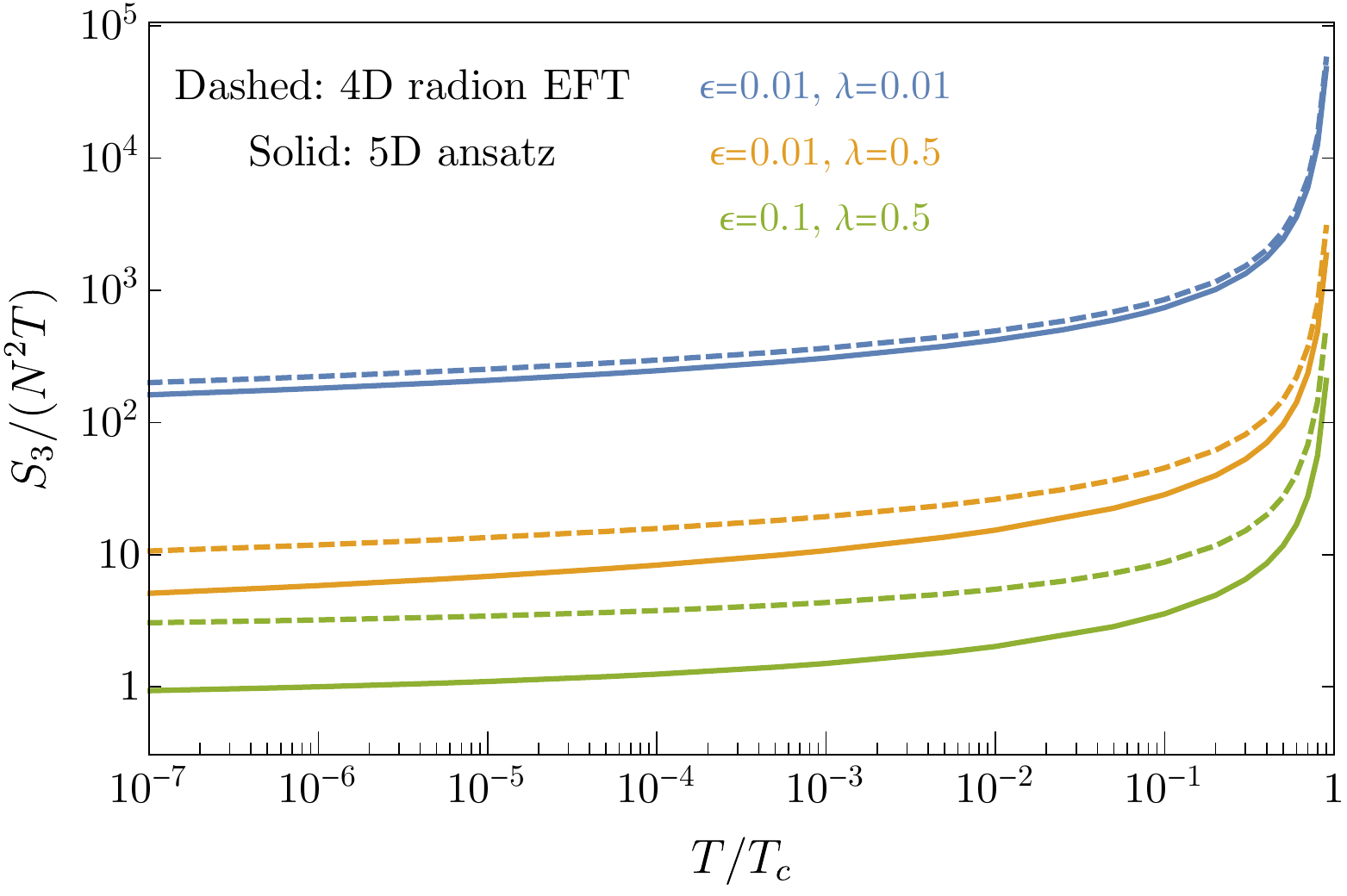}
\caption{The $O(3)$ symmetric bounce action $S_3/(N^2T)$, where $N\equiv \sqrt{16\pi^2 M_5^3}$, as a function of temperature $T/T_c$ for different choices of $\epsilon$ and $\lambda$. The solid lines denote the results using 5D ansatz, while the dashed lines use 4D radion/dilaton EFT (upto two derivatives) to estimate the bounce. }
\label{figure:S3T}
\end{figure}

 We now study the bounce for smaller $T$, where the thin-wall approximation is not valid and our ansatz bounce can no longer be shown to be a rigorous upper bound on the true bounce action. However, our bounce ansatz smoothly and simply interpolates the two phases and should still provide a very reasonable estimate of the true bounce action. Therefore, we will continue to use eq.~\eqref{fullaction2} and the boundary conditions mentioned in and around eq.~\eqref{eq:boundary_condition} to numerically obtain the $O(3)$ symmetric bounce action $S_3$ for all temperature (see solid lines in figure~\ref{figure:S3T}). 
In figure~\ref{figure:S3T} we also show the bounce action found using the two-derivative 4D 
 EFT of the dilaton (dual to radion of the 5D theory). As we discussed in 
\cite{Agashe:2019lhy}, the bounce action is dominated by such an EFT for small $\epsilon$ and $\lambda$. This is dual to the radion two-derivative EFT dominance in our 5D ansatz for small $\epsilon$ and $\lambda$, as discussed before. To compare the two-derivative approximation  to our full 5D results in this paper, we use the radion EFT Lagrangian shown in eq.~\eqref{eq:action_radion} and follow the same strategy as in \cite{Agashe:2019lhy} to determine one of the boundary conditions as $\phi'|_{\phi= 0}=\sqrt{\pi^4T^4/3}$ and thereby solve the bounce. Note that this choice extends into the region $\phi<T$ where the two-derivative EFT clearly breaks down. However, since $T<T_c\ll \langle\phi\rangle$ for small $\epsilon$ and $\lambda$, the contribution from this uncontrolled region is parametrically small and can be viewed as a subleading correction to the true bounce. 
As shown in figure~\ref{figure:S3T}, the bounce action $S_3$ calculated by the 4D
two-derivative EFT and the full 5D ansatz agree for small $\epsilon$ and $\lambda$, as expected from the argument in \cite{Agashe:2019lhy}. 
On the other hand, for larger $\lambda$ ($\lambda=0.5$) we see from figure~\ref{figure:S3T} that the two-derivative approximation is only very crude and the full 5D treatment is required. 
 
Having considered the general structure and parametrics of the PT dynamics, we now consider values of $\epsilon$ and $\lambda$ such that the observed Planck-Weak hierarchy and a successful PT are achieved. This will necessitate values of $\epsilon$ and $\lambda$ large enough that the two-derivative radion dominance approximation is insufficient and a fully 5D treatment is necessary. The 5D bounce is at least mathematically soluble in principle, qualitatively in the class described above in figure~\ref{figure:Ansatz_topology}, but in this paper we will proceed with our ansatz \eqref{fullaction2}.

\begin{figure}[h]
\centering
\includegraphics[width=0.75\linewidth]{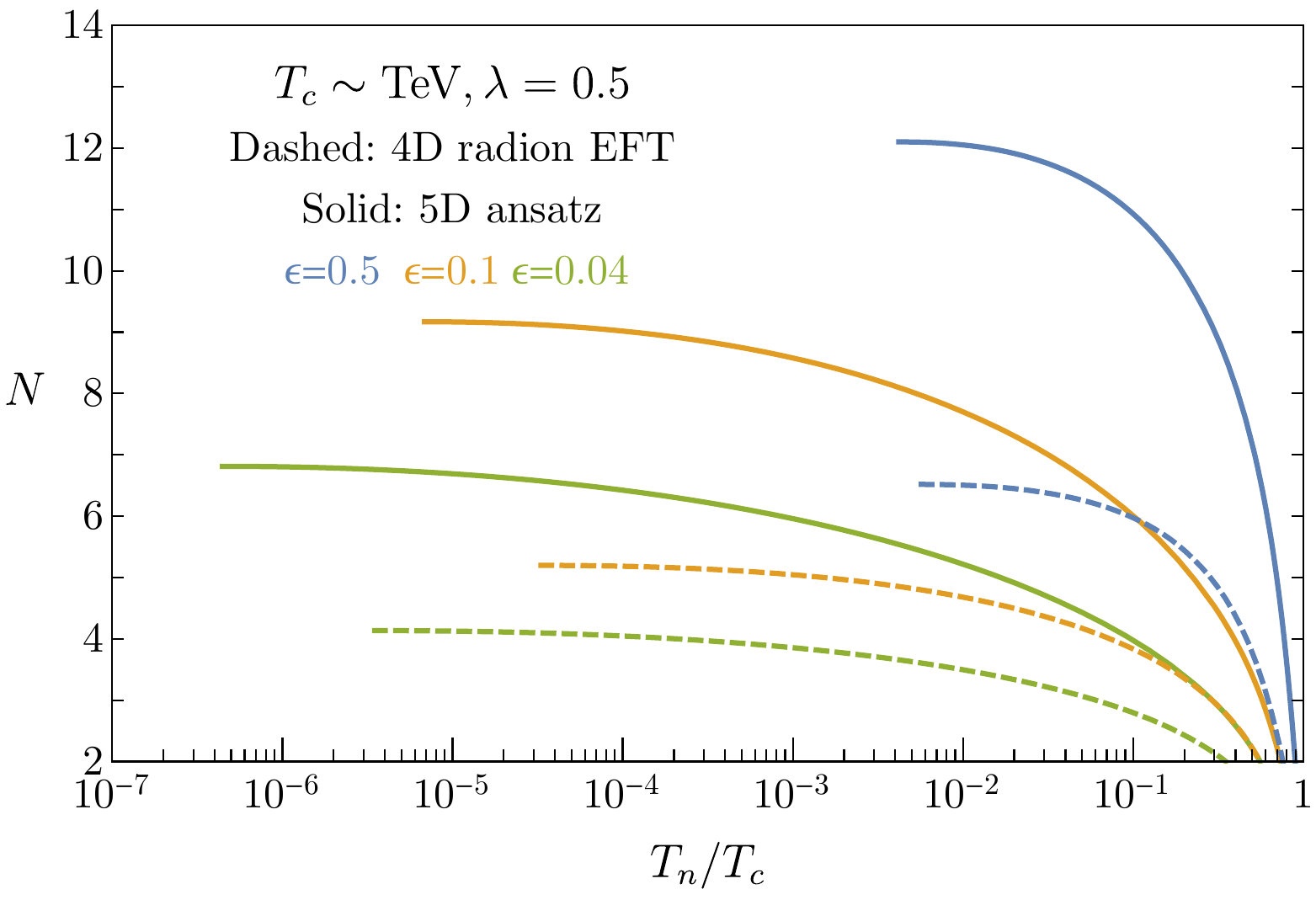}
\caption{Nucleation temperature $T_n/T_c$ (for fixed $T_c\sim \mathcal{O}(\rm TeV)$) as a function of $N$ for various $\epsilon$ for fixed $\lambda=0.5$. The solid lines denote the results using 5D ansatz, while the dashed lines use 4D radion EFT (upto two derivatives) to estimate the bounce. The end point of each curve shows the minimum $T_n$ and maximum $N$ for a given parameter choice.} 
\label{figure:nt}
\end{figure}

In an expanding universe the PT completes at a temperature $T_n$, where the nucleation rate $\Gamma(T_n)$ becomes as large as $H^4$, that is $T_n$ is found by solving $\Gamma(T_n)=H(T_n)^4$. The solid lines in figure~\ref{figure:nt} show the relation between $T_n/T_c$ and $N$ for fixed $T_c\sim \mathcal{O}(\rm TeV)$ obtained using our 5D ansatz bounce. 
 For comparison, similar curves obtained using 4D radion EFT are shown by the dashed lines in figure~\ref{figure:nt}.
 As shown in figure~\ref{figure:nt}, for a fixed $\lambda$, both the maximum $N$ and the minimum $T$ for which the PT could happen increase as $\epsilon$ gets larger. Also, for a given nucleation temperature $T_n$, as $\epsilon$ increases, completion of PT becomes possible for larger $N$ and thus in better 5D perturbative control. 

We emphasize the significance of supercooling on cosmological (dark) matter abundances \cite{Konstandin:2011dr}. For concreteness, we focus on the baryon/lepton asymmetry. We denote this asymmetry before the temperature falls to $T_c$ by $Y_{\rm before}$, where $Y\equiv \frac{n_{B}-n_{\bar{B}}}{s}$ 
 where $n_B$ ($n_{\bar{B}}$) is the number density of baryons (antibaryons). Before the PT, as the universe keeps expanding $n_B$, $n_{\bar{B}}$ and $s$ get diluted $\propto T^3$ and hence $Y$ stays constant. After the PT completes at $T=T_n$, the universes gets reheated to $T\sim T_c$ and unless the PT itself generates an asymmetry, $n_B-n_{\bar{B}}$ does not increase, while $s$ increases to $s\sim T_c^3$ and so after the PT, $Y_{\rm after}\sim Y_{\rm before} \left(\frac{T_n}{T_c}\right)^3$. 
For the observed $Y\sim 10^{-10}$, if $T_n/T_c < \mathcal{O}(10^{-3})$, then even if an $\mathcal{O}(1)$ asymmetry is generated before the PT, it would get diluted to a value below the observed asymmetry by the PT. So such a degree of supercooling, typical of the minimal $\epsilon\sim 1/25$ scenario, is inconsistent with a purely high-scale mechanism for baryogenesis. However, within our two-FP model and by choosing a larger $\epsilon \lesssim 1$, the above dilution is much smaller, making the PT compatible with baryogenesis above $T_c$. 
Of course, it is possible that baryogenesis/dark matter production may take place during or after the transition, see e.g. \cite{Konstandin:2011ds,Servant:2014bla,Matsedonskyi:2020mlz,Baldes:2020kam}, in which case such production does not get overly diluted by supercooling.


\section{Gravitational Wave Signature} \label{sec:GrWaves}
When the first order PT, described above, takes place, a stochastic background of gravitational waves (GW) gets generated. Along with the collisions of the bubbles of the confined phase, 
both the turbulence and the sound waves in the plasma, formed after bubble collisions, can source GW (for reviews see \cite{Caprini:2015zlo,Caprini:2019egz} and references therein). The properties of GW sourced by the sound waves and turbulence is an active area of research. On the other hand, the contribution from bubble collisions is analytically better understood within the so-called ``envelope approximation'', but there can be significant corrections such as discussed in \cite{Jinno:2017fby,Konstandin:2017sat,Lewicki:2020jiv,Hoeche:2020rsg,Ellis:2020nnr}.
In the following we focus on only the contribution due to bubble collisions and for simplicity use the envelope approximation to get a crude sense of the GW signal strength as a function of our parameters. It is worth keeping in mind that for some parameter space (especially for $T_n\lesssim T_c$ away from the extreme supercooling), the contribution from sound waves and turbulence can dominate over that from bubble collisions.

GW signals from bubble collisions can be characterized by the fractional abundance $\Omega_{\rm GW,b}h^2$ and peak frequency $f_p$ of GW \cite{Caprini:2015zlo}:
\begin{align}\label{eq:GW_abundance_freq}
\Omega_{\rm GW,b}h^2(f)=&1.67\times10^{-5}\left(\frac{H_{\rm PT}}{\beta_{\rm GW}}\right)^2 \left(\frac{\kappa_b \alpha}{1+\alpha}\right)^2\left(\frac{100}{g_*}\right)^{1/3}\frac{0.11 v_w^3}{0.42+v_w^2}\frac{3.8(f/f_p)^{2.8}}{1+2.8(f/f_p)^{3.8}}\\
f_p=&1.65\times10^{-4}~\text{Hz}~\frac{0.62}{1.8-0.1v_w+v_w^2}\frac{\beta_{\rm GW}}{H_{\rm PT}}\frac{T_*}{1~\rm TeV}\left(\frac{g_*}{100}\right)^{1/6}, 
\end{align}
where we assume the bubble wall velocity $v_w=1$; effective degrees of freedom $g_*=100$; that almost all of the latent heat is transferred to the bubble wall $\kappa_b\approx 1$. $\alpha$ denotes the ratio of the latent heat released in the PT to the energy in the surrounding radiation bath, which is typically $\alpha\gg 1$ for a supercooled PT. Moreover, the duration of phase transition is defined as $1/\beta_{\rm GW}$ and $H_{\rm PT}$ is the Hubble parameter during the PT. $T_*$ is the temperature of the radiation bath right after the PT.

As shown in eq.~\eqref{eq:GW_abundance_freq}, the strength and peak frequency of the gravitational wave signal produced by the phase transition depend strongly on the duration of phase transition, $1/\beta_{\rm GW}$.  $\beta_{\rm GW}$ is defined as
\beq 
\frac{\beta_{\rm GW}}{H_{\rm PT}}\equiv-\frac{T}{\Gamma} \frac{d\Gamma}{dT}\Big|_{T_n}\approx -4+T\frac{dS_b}{dT}\Big|_{T_n},
\eeq
where eq.~\eqref{Gamma} was used.

\begin{figure}
    \centering
    \includegraphics[width=0.48\linewidth]{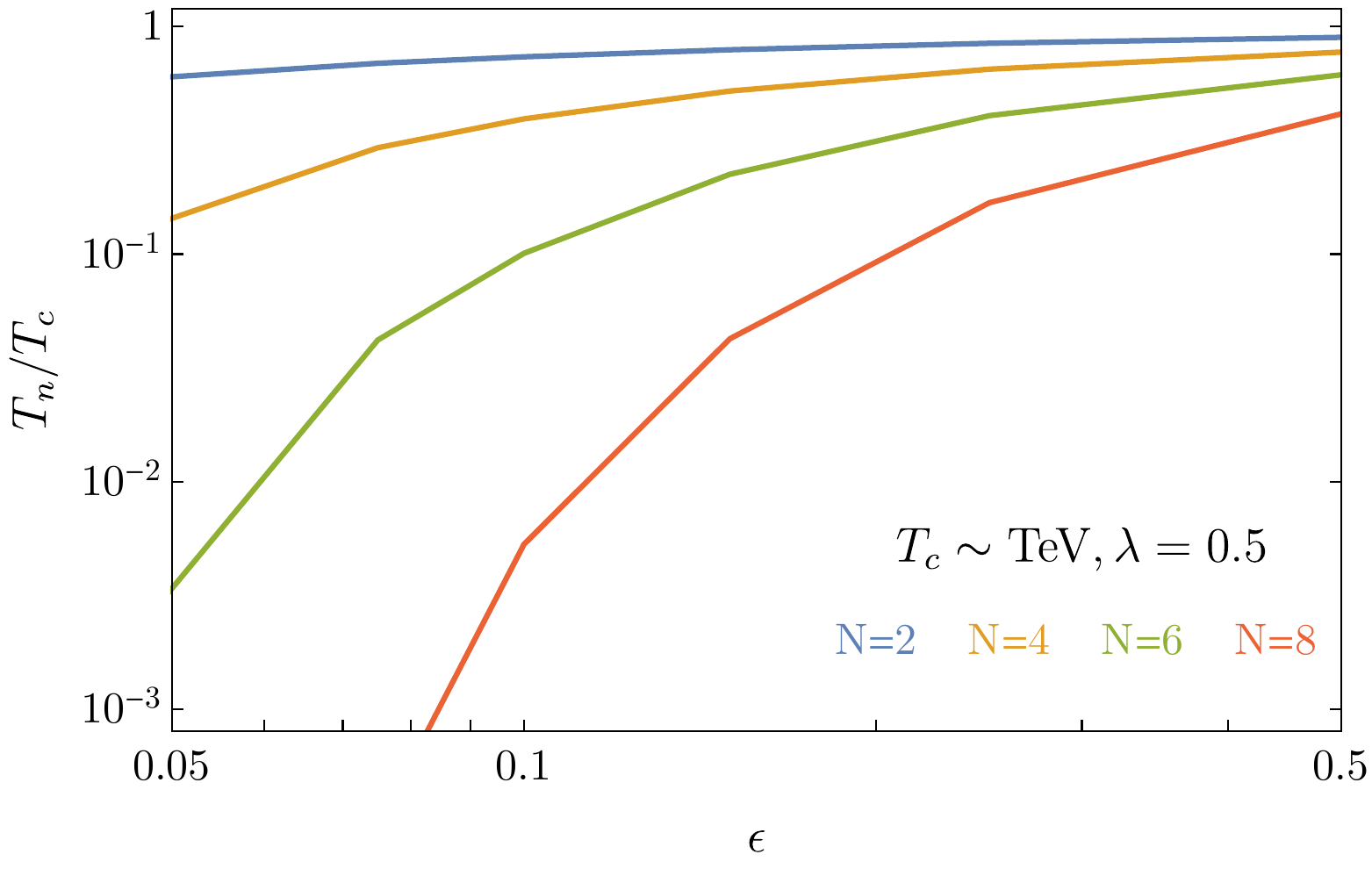}
       \includegraphics[width=0.47\linewidth]{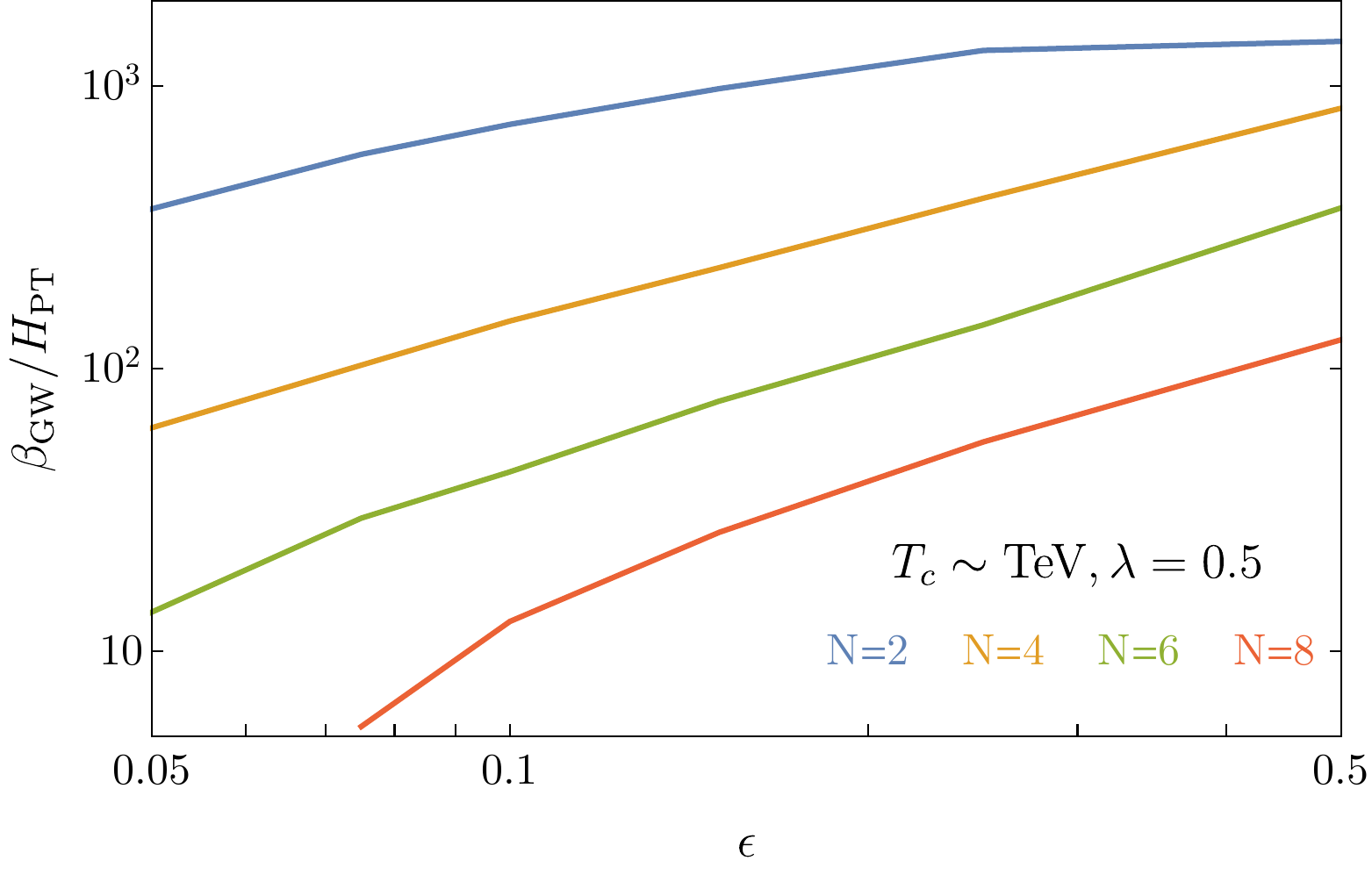}
    \caption{Nucleation temperature $T_n/T_c$ (left panel) and $\beta_{\text{GW}}/H_{\text{PT}}$ (right panel) as a function of $\epsilon$ for different choices of $N$ and  fixed $\lambda=0.5$, $T_c\sim \mathcal{O}(\text{TeV})$.}
    \label{figure:beta-n_ep}
\end{figure}

\begin{figure}
    \centering
    \includegraphics[width=0.75\linewidth]{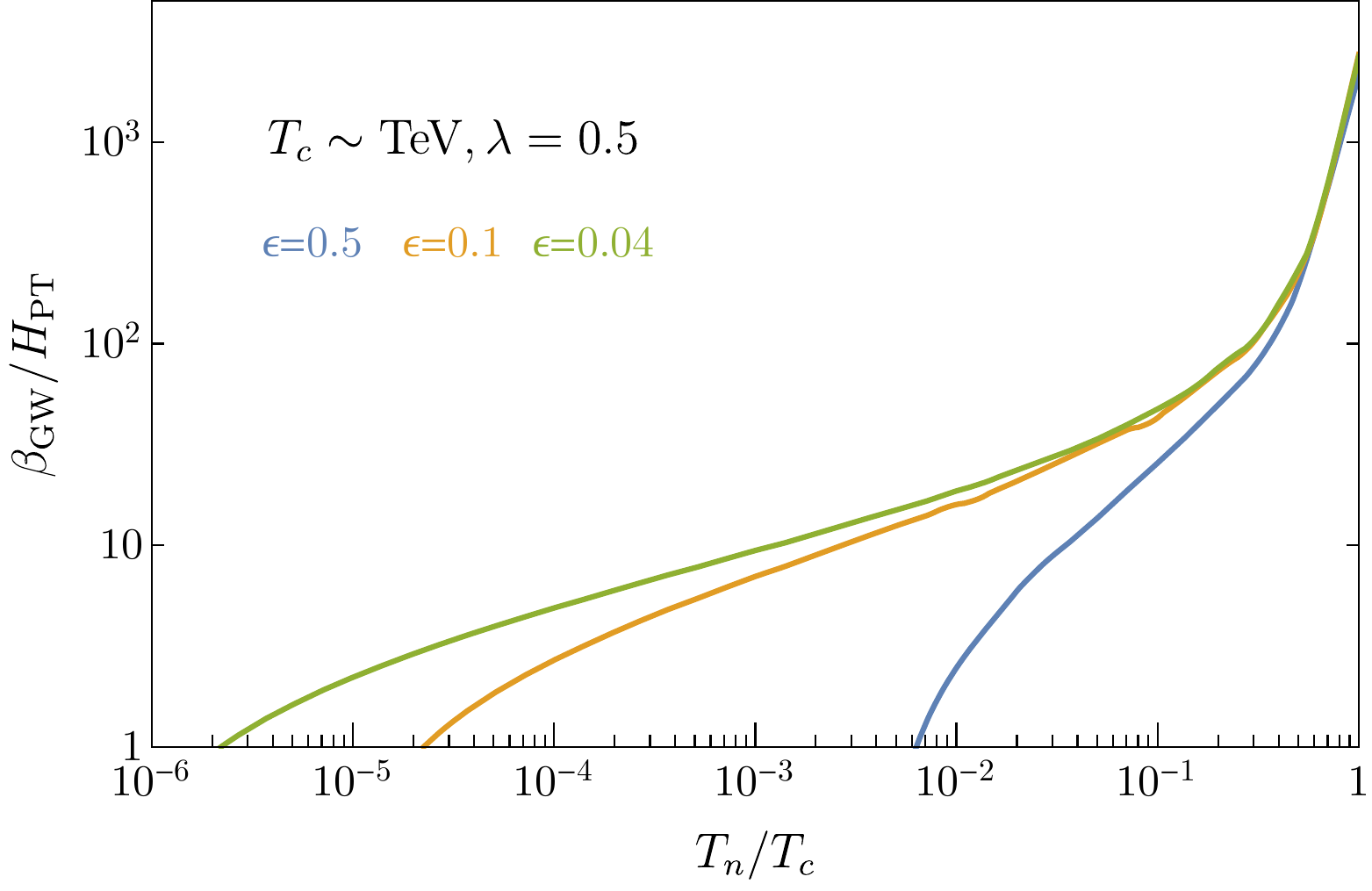}
    \caption{$\beta_{\text{GW}}/H_{\text{PT}}$ as a function of nucleation temperature $T_n/T_c$ for different choices of $\epsilon$ and  fixed $\lambda=0.5$, $T_c\sim \mathcal{O}(\text{TeV})$. On each curve, $N$ is varied while $\epsilon$ and $\lambda$ are held fixed.}
    \label{figure:beta-tn}
\end{figure}

In \cite{Agashe:2019lhy} we argued, using radion dominance approximation that in the supercooled regime $\beta_{\rm GW}$ is small for small $\epsilon$, as it was pointed out in ref.~\cite{Konstandin:2011dr}. Here we compute $\beta_{\rm GW}$ using our bounce ansatz and the results are shown in figures~\ref{figure:beta-n_ep} and~\ref{figure:beta-tn}. In figure~\ref{figure:beta-n_ep} we show the dependence of $T_n/T_c$ and $\beta_{\rm GW}$ on $\epsilon$ explicitly with different choices of $N$ and fixed $\lambda$. As we can see in (the right panel of) figure~\ref{figure:beta-n_ep}, $\beta_{\rm GW}$ gets smaller as $\ep$ decreases.
In figure~\ref{figure:beta-tn} we show  $\beta_{\rm GW}$ as a function of $T_n/T_c$ for fixed $\lambda$ and different choices of $\epsilon$ based on our 5D ansatz. It is clear from figure~\ref{figure:beta-tn} that $\beta_{\rm GW}$ drops as $T_n$ becomes smaller. One can also see from figure~\ref{figure:beta-tn} that 
to achieve a certain GW signal strength, meaning a given choice of $\beta_{\rm GW}/H_{\text{PT}}$, a theory with larger $\epsilon$ will have larger $T_n/T_c$ and thus less dilution of primordial matter abundances (see discussion in section \ref{sec:PT_not_thin_wall}). Moreover, larger $\epsilon$ also leads to larger $N$ and better perturbative control.

\begin{figure}
    \centering
    \includegraphics[width=0.8\linewidth]{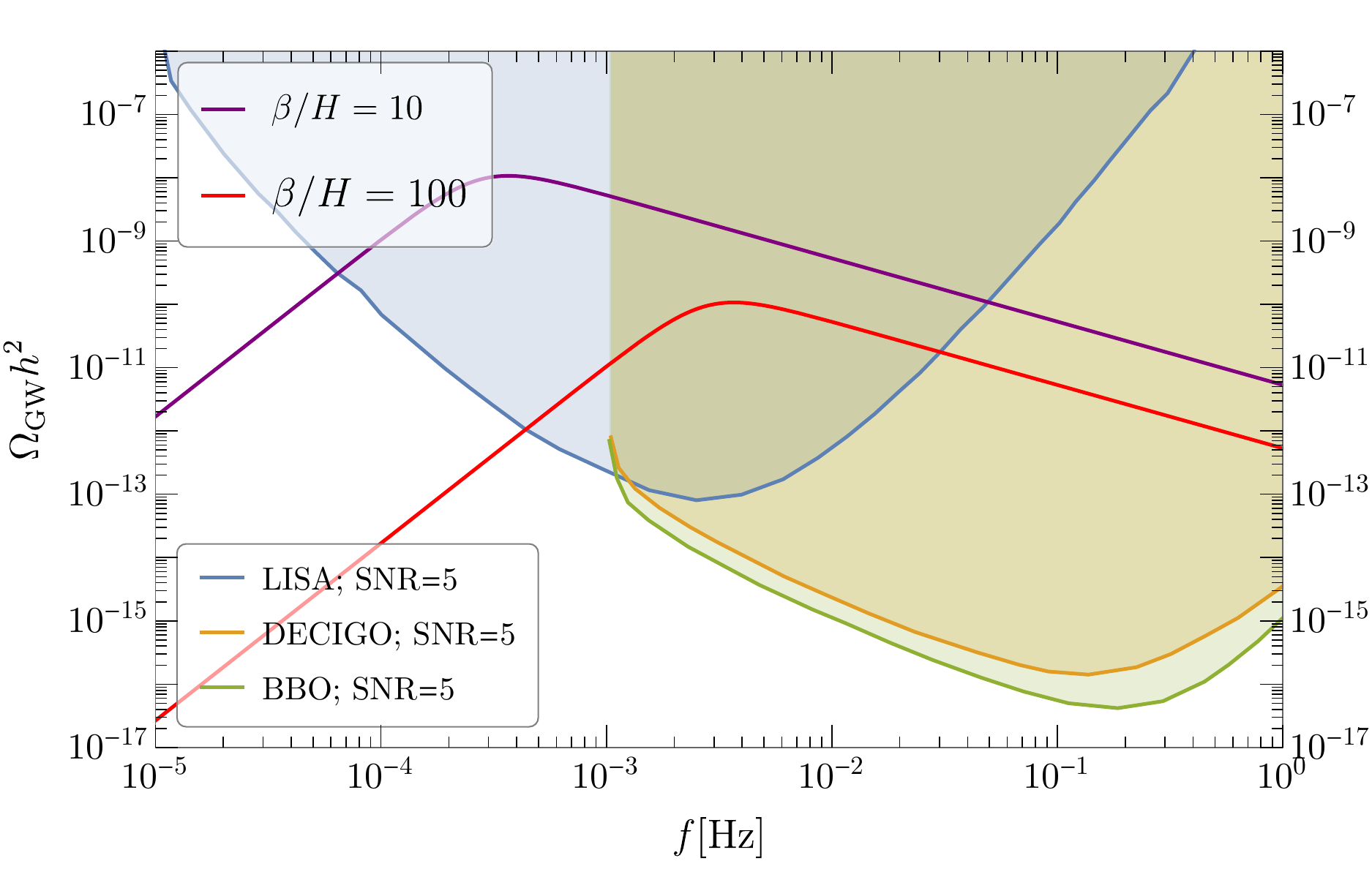}
    \caption{The spectrum of GW abundance $\Omega_{\rm GW}h^2$ as a function of GW frequency $f$ from bubble collisions. We choose two sets of benchmark parameters ($\beta/H=10$, $T_c=$ TeV) and ($\beta/H=100$, $T_c=$ TeV). The projected sensitivity of LISA, DECIGO and BBO experiments at Signal-to-Noise (SNR) $=5$ are also included.}
    \label{figure:gwsens}
\end{figure}

In figure~\ref{figure:gwsens} we show the spectrum of the fractional abundance of the GW signal for two choices of $\beta/H$ and $T_c$, considering only the more well understood contribution of bubble collisions. As mentioned before, the universe will reheat back to temperature around $T_c$ after the PT and thus we take $T_*=T_c$ in figure~\ref{figure:gwsens}.
We see that both cases can be observed by LISA, DECIGO and BBO even with this conservative estimate for the gravitational signal. For experimental sensitivity curves, we refer the reader to \cite{Schmitz:2020syl} and references therein. Although the two choices of $\beta/H$ or $T_c$ above may be realized in the standard RS models, our two-FP model allows larger values of $\epsilon$ and thus less dilution of primordial abundances and better perturbative control as mentioned before.

\section{Conclusions} \label{sec:conclusions}
Confinement-deconfinement phase transitions (PT) are interesting both from a theoretical and phenomenological point of view. This is especially true in the context of composite Higgs theories where there can be correlated future collider and gravitational wave signals. In general the confining dynamics are strongly coupled and non-perturbative, and the PT is difficult to formulate and analyze theoretically.
However, here we revisited theories having a weakly coupled holographic dual description of RS1-type, in which we are able to make progress.

We have shown how \textit{smooth} EFT-controlled 5D bounce configurations can interpolate between the two phases, which in 5D terms are given by a black-brane (dual to deconfined) phase and an IR brane (dual to confined) phase.
Our 5D construction goes beyond the conventional ansatz based on 4D radion EFT that is usually employed in the literature. While the usual radion-dominance ansatz can give a correct estimate for the rate of the PT for parametrically small values of model parameters (such as $\epsilon,\lambda$ defined in section~\ref{sec:equilibrium}), we find the 5D ansatz does become important for larger values relevant for a realistic PT consistent with observed Planck-Weak hierarchy. 

Previous studies have shown that for realistic values of model parameters, consistent with Planck-Weak hierarchy, the Universe often supercools significantly below the critical temperature $\sim$TeV, thereby diluting
primordial matter abundances generated before the PT. This makes high-scale mechanisms of (dark) matter genesis potentially incompatible with the minimal Goldberger-Wise radius stabilization mechanism (dual to the composite Higgs theories in the vicinity of a single fixed point).
However, we showed that this conclusion can easily be avoided with the simple generalization of the Goldberger-Wise bulk potential. This is dual to composite Higgs models controlled by separate UV and IR fixed points with separate critical exponents controlling particle hierarchies and the phase transition, as we proposed in ref.~\cite{Agashe:2019lhy}. Consequently, we have opened up a novel parameter space with only modest cooling, with the associated gravitational waves signal still be readily observable at future detectors, such as LISA, DECIGO and BBO. In parts of the parameter space, the stochastic gravitational wave background can be sufficiently strong that even primordial anisotropies may be observable \cite{Geller:2018mwu}.

There remain several interesting future directions. While our 5D formulation allows the bounce configuration dominating the transition rate to be semi-classically determined in principle, here we introduced a simple, qualitatively correct, bounce ansatz. We showed this implied a rigorous lower bound on the transition rate in the thin-wall regime, and a very plausible estimate more generally. However, it would be very interesting and important to obtain the true semi-classical bounce configuration. By way of inspiration, in the roughly analogous 6D model of ref.~\cite{Aharony:2005bm} (which however does not address the Hierarchy Problem), a domain wall solution was tractable and can be recast as a semi-classical bounce for the analogous phase transition. 

From a more phenomenological perspective, it would also be very interesting to develop baryogenesis mechanisms exploiting the first order nature of the phase transition. Alternately, it is interesting to consider warped phase transitions with very different critical temperatures (and thus gravitational wave frequencies) in the context of dark sectors. It is possible that the 5D holographic formulation can be useful to model aspects of the bubble and plasma dynamics relevant for a detailed understanding of the gravitational wave spectrum.

\section*{Acknowledgements}

The authors would like to thank Batoul Banihashemi, Zackaria Chacko, Theodore Jacobson, Juan Maldacena and Riccardo Rattazzi for discussions. KA, ME, SK and RS were 
supported in part by the NSF grant PHY-1914731
and by the Maryland Center for Fundamental Physics. SK was also supported in part by the NSF grant PHY-1915314 and the U.S. DOE Contract DE-AC02-05CH11231.
PD was supported in part by the NSF grant PHY-1915093.

\appendix

\section{Radion potential for the 5D model of the two fixed points}\label{app:radion_pot_5D}

In this appendix, we complete the derivation of an analytic approximation to the radion potential, eq.~\eqref{eq:radpot2FP}, in our two-FP scenario introduced in section \ref{sec:twoFP}. As discussed in section \ref{sec:twoFP}, we consider the following polynomial potential for the Goldberger-Wise field, eq.~\eqref{eq:V_GW_general}:
\beq \label{eq:PolynomPot}
V(\chi)= \frac{1}{2} m'^2 \chi^2 + \frac{1}{6} \eta \, \chi^3 +\frac{1}{24} g \, \chi^4
\eeq
where we assume the following signs for the coefficients in the potential: $m'^2<0, \eta<0$, and $g>0$. The EoM for $\chi$ is
\beq \label{eq:EoMGW2}
\frac{d^2}{d \sigma^2} \chi - 4 \frac{d}{d \sigma} \chi -V'(\chi)=0,
\eeq 
where $\sigma$ is the extra dimensional coordinate with the range $0< \sigma<  L$ and differentiation with respect to $\sigma$ will be denoted by a dot. We consider the following boundary conditions, as in eq.~\eqref{eq:BCs} for $\chi$, 
\bea
\chi(\sigma=0)=v,    \\
\dot{\chi}(\sigma=L)= \alpha,
\eea
which can be obtained from boundary potential terms for the GW field,
\bea
S_\chi^{\rm UV} && =- \int d^4x ~ \kappa \sqrt{- {\gamma}_{\rm UV}} \left( \chi^2-v^2 \right)^2,\\
S_\chi^{\rm IR} && =  \int d^4x ~  \sqrt{-{\gamma}_{\rm IR}} ~  \alpha \chi,
\eea
where $S^{\rm UV}_\chi$ and $S^{\rm IR}_\chi$ denote the UV and IR brane terms respectively. To impose the UV boundary condition, $\chi(\sigma=0)=v$ , one needs to take the limit of large $\kappa$. In this limit the only effect of this term is setting the boundary condition and does not have any extra contribution to the radion potential. The $S^{\rm IR}_\chi$ term, on the other hand, will contribute to the radion potential.

As mentioned in section~\ref{subsec:twoFP_5D}, the above potential in eq.~\eqref{eq:PolynomPot} has two extrema\footnote{The potential of eq.~\eqref{eq:PolynomPot} has another extremum in $\chi<0$, but that is not important for the solutions for $\chi$ that we consider here, as we choose the UV boundary value $v$ to be on the range $0<v<\chi_*$.} at
\begin{align}
    \chi=0~\textrm{and}~\chi=\chi_*=\frac{3}{g}\left(-\frac{\eta}{2}+\sqrt{\frac{\eta^2}{4}-\frac{2gm^{\prime 2}}{3}}\right),
\end{align} 
corresponding to two constant-$\chi$ solutions of EoM.  To compute the radion potential analytically, we approximate the Goldberger-Wise bulk potential by a piecewise-quadratic potential given by
\bea\label{eq:Vchi_piecewise}
\tilde{V}_\chi(\chi)=
\begin{cases}
     \frac{1}{2} m'^2 \chi^2  & \chi\leq \chi_{\rm m}\\
     \frac{1}{2} m^2 (\chi-\chi_*)^2-C
     & \chi>\chi_{\rm m} 
\end{cases} ,
\eea
where $m^{ 2} \equiv V_\chi^{''}(\chi_*)=-2m^{\prime 2}-\eta \chi_*/2$ and we have matched the two approximations at $\chi_{\rm m}=(-\eta+\sqrt{\eta^2-2gm^{\prime 2}})/g$ corresponding to the inflection point of the potential in eq.~\eqref{eq:PolynomPot}. The constant $C= \frac{1}{2}m^{ 2} (\chi_{\rm m}-\chi_*)^2-\frac{1}{2} m^{\prime 2} \chi_{\rm m}^2$ ensures the continuity of $\tilde{V}_\chi(\chi)$. We are interested in the case of $|m'^2| \ll m^2$, using which we get $\chi_{\rm m}\approx\frac{2}{3}\chi_*$ and $C\approx\frac{1}{18}m^2\chi_*^2$.

We choose $v$ in the range $0<v<\chi_*$, and close to $\chi=0$ . With this choice, for small enough $L$, the field value is always smaller than $\chi_{\rm m}$ and thus $\tilde{V}_\chi(\chi)$ in eq.~\eqref{eq:Vchi_piecewise} reduces to the standard quadratic potential.  In this case, the standard Goldberger-Wise solution applies \cite{Chacko:2013dra}:
\bea\label{eq:standard_chi}
\chi(\sigma)\approx v e^{\epsilon'\sigma}+\frac{\alpha}{4}e^{(4-\epsilon')(\sigma-L)}~~~~~(\textrm{for}~L<L_{\rm m}),
\eea
where $\epsilon'=2-\sqrt{4+m'^2}$ and we have dropped terms that are higher order in $\epsilon'$ and $e^{-L}$.
We define $L_{\rm m}$ as the special value of $L$ that satisfies:
\bea \label{eq:defineLm}
\chi(L)|_{L=L_{\rm m}}=\chi_{\rm m}&\rightarrow & L_{\rm m} \approx \frac{1}{\epsilon'}\ln\left(\frac{\chi_{\rm m}-\alpha/4}{ v }\right).
\eea
It is easy to see from eq.~\eqref{eq:standard_chi} that in the region $L<L_{\rm m}$, the field profile $\chi(\sigma)$ always stays smaller than $\chi_{\rm m}$ and thus it was self-consistent to use eq. \eqref{eq:standard_chi}.

For larger values of $L$, i.e. for $L>L_{\rm m}$, the solution has the form 
\bea\label{eq:chi_piecewise}
\chi(\sigma)=
\begin{cases}
A_1 e^{\epsilon' \sigma}+ A_2 e^{(4-\epsilon') \sigma}   & \sigma<\sigma_{\rm m}    \\
B_1 e^{-\epsilon (\sigma-\sigma_{\rm m})}+B_2 e^{(4+\epsilon) (\sigma-\sigma_{\rm m})} +\chi_*  & \sigma>\sigma_{\rm m}
\end{cases}
\eea
in which $\sigma_{\rm m}$ is defined as $\chi(\sigma_{\rm m})=\chi_{\rm m}$ and $\epsilon=-2+\sqrt{4+m^2}$.\footnote{We choose to work with positive $\epsilon$ and $\epsilon'$ and thus there is a sign difference in their expressions.}
In the expression above $A_i$, $B_i$, $\sigma_{\rm m}$ are independent of $\sigma$, but are in general $L$-dependent.  In addition to the two boundary conditions, there are three other conditions that we have to impose to solve for the five unknowns $A_{1,2}$, $B_{1,2}$ and $\sigma_{\rm m}$. These conditions are continuity of $\chi$ and $\dot{\chi}$ at $\sigma_{\rm m}$, implied by continuity of $V(\chi)$(or finiteness of $V'(\chi)$) and the EoM for $\chi$, and that the field value is equal to $\chi_{\rm m}$ at $\sigma_{\rm m}$. So we need to solve the following set of equations:
\bea \label{eq:setofeqtoy2}
\begin{cases}
A_1+A_2=v  \\
A_1 e^{\epsilon' \sigma_{\rm m} } + A_2 e^{(4-\epsilon') \sigma_{\rm m}}=\chi_{\rm m}    \\
\epsilon' A_1 e^{\epsilon' \sigma_{\rm m} } + (4-\epsilon') A_2 e^{(4-\epsilon') \sigma_{\rm m}}= -\epsilon B_1+ (4+\epsilon) B_2 \\
B_1+B_2= \chi_{\rm m}-\chi_* \\
-\epsilon B_1 e^{-\epsilon (L- \sigma_{\rm m}) } +(4+\epsilon) B_2 e^{(4+\epsilon) (L- \sigma_{\rm m})}=\alpha
\end{cases}
\eea    
In terms of $\sigma_{\rm m}$ the first two and last two equations can be solved separately:
\bea
\begin{cases}
A_1  \approx v -\chi_{\rm m}e^{-(4-\epsilon')\sigma_{\rm m}} \\
A_2  \approx \chi_{\rm m} e^{-(4-\epsilon')\sigma_{\rm m}}-v e^{-(4-2 \epsilon') \sigma_{\rm m}} \\
B_1  = \frac{(4+\epsilon) (\chi_{\rm m}-\chi_*) e^{( 4 + \epsilon ) (L-\sigma_{\rm m})}-\alpha }{( 4 + \epsilon ) e^{(4 + \epsilon) (L-\sigma_{\rm m})}+ \epsilon \, e^{-\epsilon (L-\sigma_{\rm m})} } \approx  (\chi_{\rm m}-\chi_*) -\frac{\alpha}{4+\epsilon}e^{- ( 4 + \epsilon) (L-\sigma_{\rm m})}               \\
B_2= \frac{\alpha- \epsilon (\chi_*-\chi_{\rm m})  \, e^{-\epsilon (L- \sigma_{\rm m})}}{( 4 + \epsilon ) e^{( 4 + \epsilon ) (L-\sigma_{\rm m})}- \epsilon \, e^{-\epsilon (L-\sigma_{\rm m})}} \approx \frac{\alpha}{4+\epsilon}e^{- ( 4 + \epsilon ) (L-\sigma_{\rm m})} -\frac{\epsilon}{4+\epsilon}(\chi_*-\chi_{\rm m})e^{- ( 4 + 2 \epsilon ) (L-\sigma_{\rm m})}
\end{cases}
\eea
Now putting these into the third equation of eqs.~(\ref{eq:setofeqtoy2}), we find an equation for $\sigma_{\rm m}$:
\bea
-(4-2\ep') v e^{\epsilon' \sigma_{\rm m}}+(4-\ep') \chi_{\rm m} \approx \epsilon(\chi_*-\chi_{\rm m})+\left( \alpha-\epsilon(\chi_*- \chi_{\rm m})  e^{-\epsilon ( L- \sigma_{\rm m} )} \right) e^{- ( 4 + \epsilon ) (L-\sigma_{\rm m})}.\nonumber\\
\eea
The right-hand-side of this equation becomes small quickly as $L- \sigma_{\rm m}$ increases, making $e^{\epsilon' \sigma_{\rm m}}$ weakly dependent on $L$. Expanding in  $\epsilon'$ and $e^{-(L-\sigma_{\rm m})}$ we get
\beq
e^{\epsilon' \sigma_{\rm m}} \approx \frac{\chi_{\rm m}-\frac{\ep}{4} (\chi_*-\chi_{\rm m})+\frac{\ep'}{4}\chi_{\rm m}}{v}-\frac{\alpha}{4 v}\left( \frac{\chi_{\rm m}-\frac{\ep}{4} (\chi_*-\chi_{\rm m})+\frac{\ep'}{4}\chi_{\rm m}}{v} \right)^\frac{4+\ep}{\ep'}  \phi^{4+\epsilon}.
\eeq

We now move on to calculate the radion potential. As mentioned in section~\ref{sec:equilibrium}, radion action is obtained by promoting $e^{-L}$ to $\phi(x)$ in $S_{\rm 5D}$ (eq.~\eqref{eq:RSaction}). The dominant contribution to 4D radion potential $V_{\rm rad}(\phi)$ comes from $S_\chi$ (eq.~\eqref{eq:GWaction}) after integrating over the 5th dimension.
For $\phi>\phi_{\rm m}\equiv e^{-L_{\rm m}}$, the Goldberger-wise scalar $\chi(\sigma)$ has the standard form eq.~\eqref{eq:standard_chi} and thus the radion potential is the same as eq.~\eqref{radionpot1}. For $\phi<\phi_{\rm m}$, Goldberger-wise scalar $\chi(\sigma)$ has the form in eq.~\eqref{eq:chi_piecewise} given our piecewise-potential approximation in eq.~\eqref{eq:Vchi_piecewise}. Similar to case of free Goldberger-Wise field, we can use the EoM  in the bulk and then integrate by parts to get the radion potential in terms of only the boundary terms:
\bea
V_{\rm rad}(L)\supset \left[\frac{1}{2} e^{-4 \sigma} \chi \dot{\chi} \right ]_{\sigma=0}^{\sigma_{\rm m}} + \left[\frac{1}{2} e^{-4 \sigma} (\chi-\chi_*) \dot{\chi} \right ]_{\sigma=\sigma_{\rm m}}^{L}-e^{-4 L} \alpha \chi (L) +\frac{C}{4}\left[ e^{-4\sigma} \right]_{\sigma=\sigma_{\rm m}}^{L},\nonumber\\ 
\eea
which in terms of $\phi$ and the coefficients $A_{1,2}$ and $B_{1,2}$ becomes
\bea\label{eq:Vrad_int}
V_{\rm rad}(\phi) &\supset - \frac{1}{2}\alpha \phi^4 \left(  2 \chi_*  + B_1 \left( e^{\sigma_{\rm m}} \phi  \right)^{\epsilon} +B_2 \left( e^{ \sigma_{\rm m}} \phi \right)^{-(4+\epsilon)}  \right)   +\frac{1}{2} \chi_* e^{-4 \sigma_{\rm m}}  \left(  -\epsilon B_1 +(4+\epsilon) B_2 \right)\nonumber \\
 & - \frac{1}{2}v \left(  \epsilon' A_1 +(4-\epsilon') A_2 \right)-\frac{C}{4}(e^{-4\sigma_{\rm m}}-e^{-4 L}).   
\eea
Substituting the coefficients in eq.~\eqref{eq:Vrad_int} and
including the IR brane tension detuning and dropping the constant terms, we obtain the following radion potential keeping the leading order in $\epsilon$ and $\epsilon'$: 
\bea\label{eq:radion_potential_full}
V_{\rm rad} (\phi) \approx
\begin{cases}
 \tau \phi^4  -  \alpha v \phi^{4-\epsilon'} & \phi> \phi_{\rm m} \\
(\tau-\alpha \chi_*) \phi^4 + \alpha (\chi_*-\chi_{\rm m}) \left( \frac{\chi_{\rm m}-\frac{\ep}{4} (\chi_*-\chi_{\rm m})}{v} \right)^{\epsilon/\epsilon'} \phi^{4+\epsilon} & \phi \ll \phi_{\rm m} \\
\end{cases},
\eea 
where $\tau\equiv\tau_{_{\rm IR}}+12M_5^3-\alpha^2/8$ and $\tau_{_{\rm IR}}$ is allowed to be detuned away from the RS value $-12M_5^3$.

 We can choose $\tau>0$ and $\tau'\equiv\tau - \alpha \chi_*<0$ while all other parameters in eq.~\eqref{eq:radion_potential_full} are  positive. For this choice of parameters, the above potential in eq.~\eqref{eq:radion_potential_full} has only one minimum which is located in the $\phi<\phi_{\rm m}$ range:
\beq
\langle\phi\rangle\sim \left( \frac{v}{\chi_{\rm m}(1-\ep/8)} \right)^{1/\epsilon'} \left(\frac{ -2\tau'}{ \alpha \chi_{\rm m}}\right)^{1/\epsilon},
\eeq
where we have used $\chi_{\rm m}\approx \frac{2}{3} \chi_*$ to simplify the expression.
This completes the derivation of eqs~\eqref{eq:radpot2FP} and \eqref{eq:radmin}, mentioned in section \ref{sec:twoFP}.

\bibliographystyle{JHEP}
\bibliography{RSPT5D}
\end{document}